\documentclass[useAMS,usenatbib]{mn2e}
\usepackage{graphicx}
\usepackage{dcolumn}
\usepackage{bm}
\usepackage{color}

\newcommand{\beq}{\begin{equation}}
\newcommand{\eeq}{\end{equation}}

\newcommand{\cm}{{\mathrm{cm}}}

\newcommand{\s}{\mathrm{s}}

\def\l{\lambda}
\def\pv{P_{\rm v}}

\def\pvv{P_{\vec{{\rm v}}}  }
\def\r{\rho}
\def\ro{\rho_0}

\def\Vvt{V(v_1, v_2;\, t_1, t_2)}
\def\Xvt{\xi(v_1, v_2;\, t_1, t_2)}
\def\Xmvt{\xi_1(v_1, v_2;\, t_1, t_2)}
\def\Xsvt{\xi_2(v_1, v_2;\, t_1, t_2)}
\def\p1{{\rm p}_1}
\def\psf{{\rm PSF}(p_{\rm t}-p_{\rm o})}
\def\pt{p_{\rm t}}
\def\po{p_{\rm o}}

\def\pd{{\rm p_d}}

\def\DX{\Delta \xi}
\def\D{\Delta}
\def\Erf{{\rm Erf}}

\def\PiL{P_{i,L}}

\def\vv{\vec{v}}

\def\Ohat{\hat{\Omega}}
\def\Ohatp{\hat{\Omega}_p}
\def\T{\theta}
\def\capT{\Theta}
\def\w{\omega}

\def\Fres{F_{\rm res}}
\def\deg{^\circ}

\title[Detecting unresolved moving sources in a diffuse background]{Detecting unresolved moving sources in a diffuse background}
\author[Alex Geringer-Sameth and Savvas M. Koushiappas]
{
	Alex Geringer-Sameth\thanks{email: alex\_geringer-sameth@brown.edu} and
	Savvas M. Koushiappas\thanks{email: koushiappas@brown.edu} \\
	Department of Physics, Brown University, 182 Hope Street, Providence, RI 02912
}

\begin{document}

\date{Submitted 2011 April 21}


\maketitle


\begin{abstract}
We present a statistical technique which can be used to detect the presence and properties of moving sources contributing to a diffuse background. The method is a generalization of the 2-point correlation function to include temporal as well as spatial information. We develop a formalism which allows for a derivation of the spacetime 2-point function in terms of the properties of the contributing sources. We test this technique in simulated sky maps, and demonstrate its robustness in identifying the presence of moving and stationary sources. 
Applications of this formalism to the diffuse gamma-ray background include searches for solar system bodies, fast moving primordial black holes, and dense cores of dark matter proto-halos in the solar neighborhood. Other applications include detecting the contribution of energetic neutrinos originating in the solar system, as well as probing compact objects in long-timeline lensing experiments.
\end{abstract}


\section{Introduction}

Diffuse background light is very important in understanding conditions and classes of objects in the Universe. This is due to the fact that the spectral, spatial, and amplitude information in a diffuse background is linked to the properties of the otherwise unresolved contributing sources. For example, microwave background measurements include contributions of cosmic origin \citep{2009ApJS..180..330K}, as well as foregrounds of Galactic origin \citep{Bernardi:2005hx,2011MNRAS.412.2383C,2008ApJ...680.1222D,Dobler:2008av,2009ApJS..180..265G}. As another example, $\gamma$-ray background measurements include contributions from unresolved blazars 
\citep{1993MNRAS.260L..21P, 1996ApJ...464..600S, 2000MNRAS.312..177M, 1998ApJ...496..752C,2006ApJ...643...81N, 2009RAA.....9.1205B, 2010ApJ...710.1530V, 2009MNRAS.400.2122A, 2009ApJ...702..523I, Dodelson:2009ih}, inverse Compton scattering of CMB photons by electrons accelerated at shocks around galaxy clusters and cosmic filaments \citep{2000Natur.405..156L,2002MNRAS.337..199M,1998APh.....9..227C,2007ApJ...667L...1M}, starburst galaxies \citep{2002ApJ...575L...5P}, cosmic ray interactions with atomic and molecular gas in the Milky Way \citep{1986A&A...157..223D,2009PhRvL.103y1101A}, as well as the possible annihilation of dark matter \citep{Ando:2009fp,Ando:2006cr,2007JCAP...04..013C,Cuoco:2007sh,Fornasa:2009qh,Hooper:2007be,2009JCAP...07..007L,2010ASPC..426...87S,SiegalGaskins:2009ux,SiegalGaskins:2008ge,Taoso:2008qz,2010PhRvD..82l3511B}. 

Background events may be divided into two classes. Some events are generated by localized sources while others are generated by mechanisms which cannot be localized. In the first class the sources can be either spatially fixed (in celestial coordinates) or may exhibit proper motion (i.e. over a period of time their displacements are larger than the angular resolution of the detector).

Using again the diffuse $\gamma$-ray background as an example, unresolved blazars, starburst galaxies, and emission from structure formation shocks would be considered spatially fixed sources of background. Cosmic ray events with interstellar gas would be considered a non-localized random process. Sources of background which will exhibit proper motion include the interaction of energetic cosmic rays with solar system bodies (e.g., small objects in the asteroid belt or objects in the Kuiper belt and the Oort cloud) \citep{1984JGR....8910685M,2008ApJ...681.1708M,2009ApJ...692L..54M}, dark matter annihilation around primordial black holes \citep{Mack:2006gz,2010ApJ...720L..67L}, and potentially nearby remnants of high density dark matter density peaks \citep{2006PhRvL..97s1301K,2008MNRAS.384.1627P,2008PhRvD..78j1301A,2009NJPh...11j5012K}. In all these cases, individual emission from any single object is not distinguishable, but the sum of these contributions may contribute to the diffuse $\gamma$-ray background. 

Correlations between individual events can help disentangle the contribution of various sources to the background. In this manuscript we present a formalism and a technique that can be used to identify the presence of background sources that exhibit spatial motion. In Sec.~\ref{sec:overview}  we present an overview of the problem. In Sec.~\ref{sec:definitions} we detail definitions that are used in the statistical techniques that follow. This allows us to write down the formal definition of the spacetime 2-point correlation function, which can be used to extract the moving signal in the diffuse background. In Sec.~\ref{sec:2d} we derive the form of the spacetime correlation function in 2 dimensions.  A quantitive account of the uncertainty in the method is found in Sec.~\ref{sec:errors}. In Sec.~\ref{sec:demos} we demonstrate the method's robustness in toy experiments and comment about the use of an instrumental point spread function. We generalize the formalism to realistic problems in 3 dimensions in Sec.~\ref{sec:3d}, discuss generalizations of the formalism in Sec.~\ref{sec:discussion} and discuss applications and conclude in Sec.~\ref{sec:conclusions}. 

\begin{figure*}
\centering
\resizebox{2.2in}{!}
	{\includegraphics{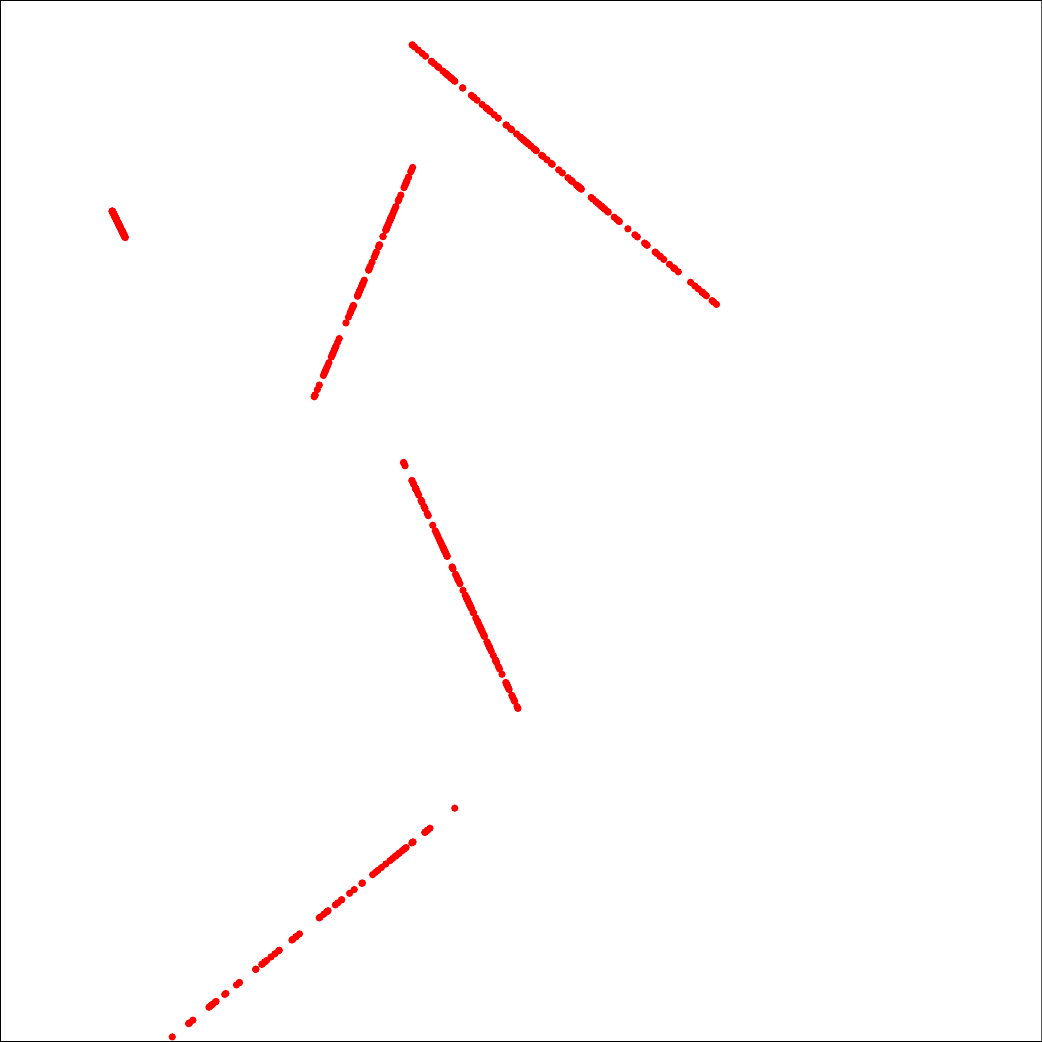}}
\resizebox{2.2in}{!}
	{\includegraphics{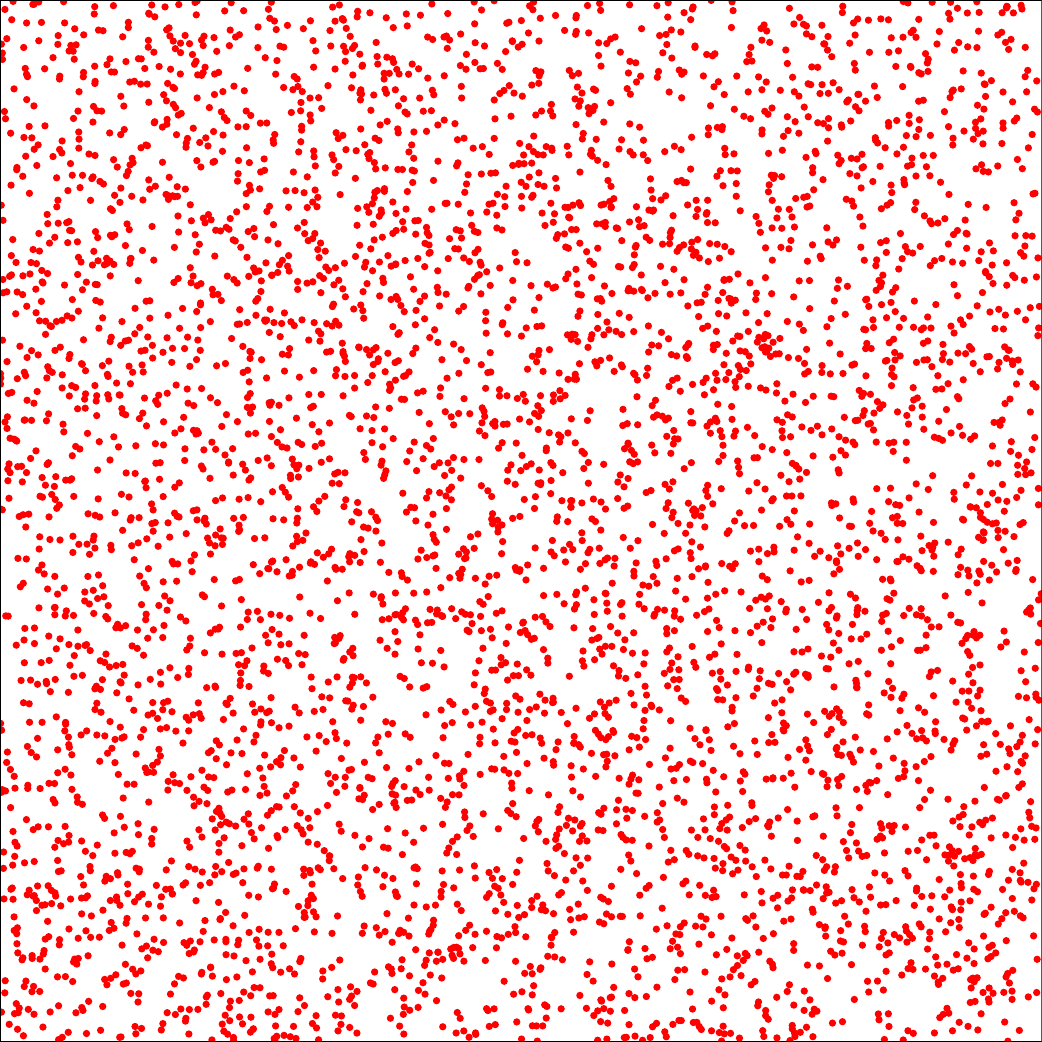}}
\resizebox{2.2in}{!}
	{\includegraphics{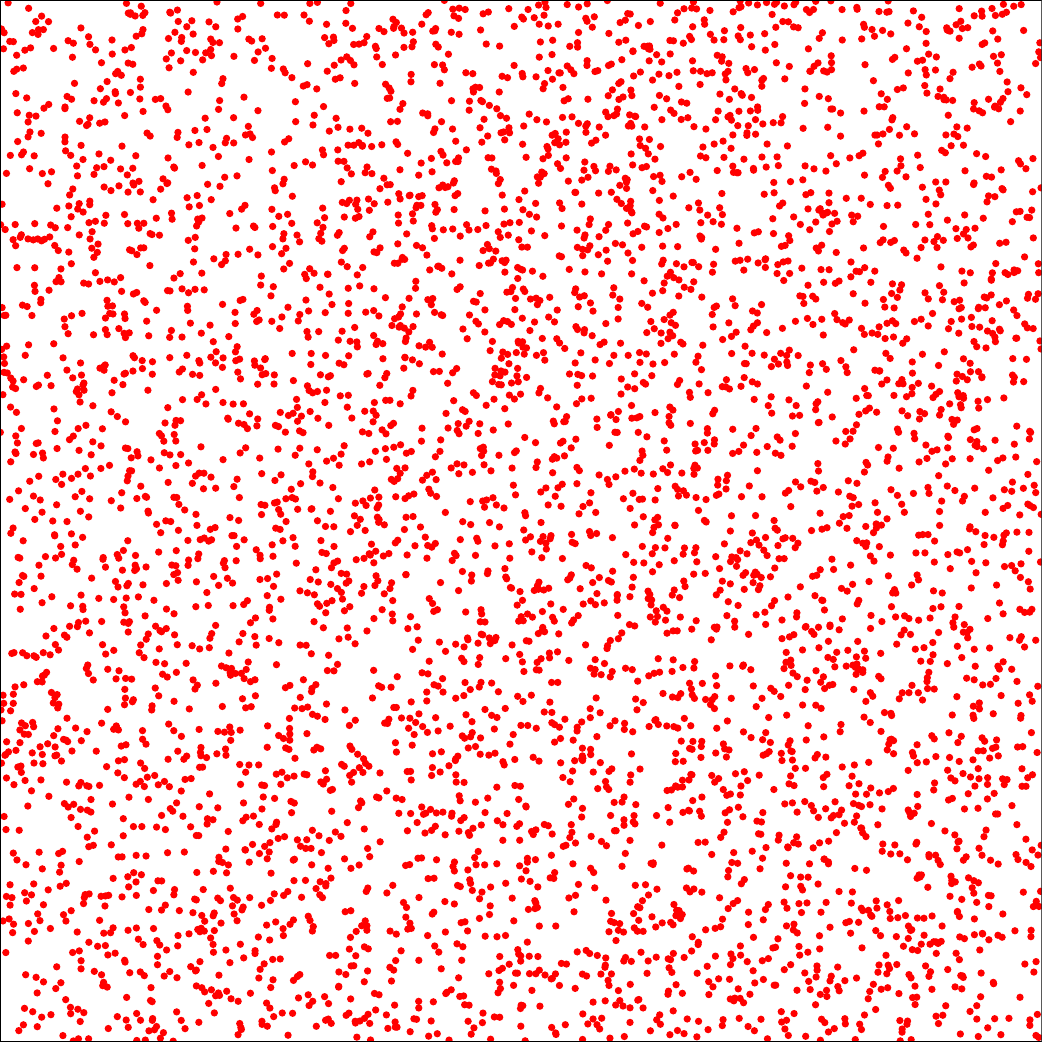}}	
\caption{Illustration of the two limits in the problem. The first figure contains 5 objects each with event rate 10 and the second contains 50,000 objects with event rate 0.01. The third figure contains the same number of events as the second but they are distributed randomly. Naively, it is impossible to tell which of the last two figures contains random events and which contains moving objects.}
\label{fig:2limits}
\end{figure*}

\section{Overview of the problem}
\label{sec:overview}

	Suppose we have some objects moving on a 2-dimensional surface, each with a constant velocity. Every so often the objects emit photons, which, when detected, we call ``events''. We record the location and time of each photon detection. The problem we are interested in is to take this collection of events and extract information about the objects: their existence, their velocity distribution, their density distribution, and their event rate or luminosity (i.e. the rate at which photons are emitted from each object). 
	
	There are two natural variables in this problem which ought to determine how difficult it will be to extract this information: the event rate of the objects and their number density. If there are very few objects and their luminosities are very high it should be easy to identify the path of each object individually. In the opposite limit the objects' luminosities are small but their number density is large. In this case it will be difficult to identify the sequences of events that trace the paths corresponding to individual objects. 
	
	These two limits are represented in Fig.~\ref{fig:2limits}. Each panel in Fig.~\ref{fig:2limits} is a plot of the location of all events in 10 arbitrary units of time\footnote{In these examples time and distance have arbitrary units and from now on these units will be set equal to $1$. A phrase like ``luminosity equal to $10$'' means an event rate of $10$ per unit time; ``an average speed of $5$'' means $5$ units of distance per unit time, etc.}. The left panel contains 5 objects each having a luminosity of 10 and an average speed of 5. The middle panel contains 50,000 objects each with an event rate of 0.01 and drawn from the same velocity distribution as before. The right panel contains the same number of events as the middle panel, but they occur at random positions and times (i.e., there are no ``objects''). In the left panel it is easy to measure the speed and event rate of every object (each generating about $100$ events in total). This task is impossible, by eye, for the middle panel where each object generates $0.1$ events on average. Indeed, it is even difficult to say whether or not the events come from objects at all, or if they are simply generated randomly as in the right panel. In practice, the left panel is analogous to resolvable sources in the absence of any contaminating backgrounds while the middle and right panels represent diffuse backgrounds in the sky. Our goal is to be able to distinguish between the middle and right panels while learning something about the objects in the middle panel.
	
	The technique we employ is an application of the 2-point correlation function. One takes every pair of events and calculates their time separation and ``velocity separation'' (their spatial separation divided by their temporal separation). One can then bin this data and make a 3-dimensional plot of number of pairs as a function of both time separation and velocity separation. The shape of this surface reveals information about the contributing objects. For instance, if all the objects are moving exactly at speed $v$, there will be lots of pairs of events whose velocity separation is $v$. The effect will be a ridge in this 2-dimensional parameter space. 

	The situation can be made more realistic. Instead of the moving objects all having speed $v$, their speeds could be drawn from a distribution. Their event rates could also be drawn from a distribution. In fact we might have many different populations of objects each having a different set of distributions for speed and luminosity. On top of this we could add a set of completely random events: a Poisson process such that there is some constant probability that an  event occurs in any small region of spacetime. Below we will systematically discuss all these possibilities. First we present the simple 2-dimensional case with one class of moving objects along with a component of random events. This is the easiest way to present our formalism. Then we straightforwardly generalize to a realistic case where a diffuse background is made up of signals coming from various populations of objects as well as random processes.

\section{Definitions} 
\label{sec:definitions}

	The analysis takes place on a 2-dimensional sky map, which is a collection of discrete signals that we define as ``events''. Each event is assigned a spatial coordinate (position) and a time coordinate. For example, in the case of the Fermi Gamma-ray Space Telescope (Fermi), discrete signals are $\gamma$-ray events recorded by the Large Area Telescope (LAT). The position is the location on the sky where the photon originated, and the time is the time of detection. 
It is important to note that in realistic experiments the data comes not as a list of (position, time) for each event but as a list of (point spread function, time) for each event. 
The analysis that follows can be reworked for this more realistic situation. However, we will start out by assuming that we simply have a collection of events where each event is specified by a position and a time.
	
	As we are interested in sources of events that can have velocities we also need a notion of distance. For realistic sky maps, the distance between two events is defined to be their angular separation. In our toy  model with objects moving on a 2-dimensional surface, the distance between events is their Euclidean distance. We also define the ``velocity separation'' between two events to be the distance between them divided by their time separation. With these definitions, the appropriate way to visualize the data is in a spacetime diagram where each event has both position and time coordinates.

\begin{figure}
\centering
\resizebox{3in}{!}
{\includegraphics{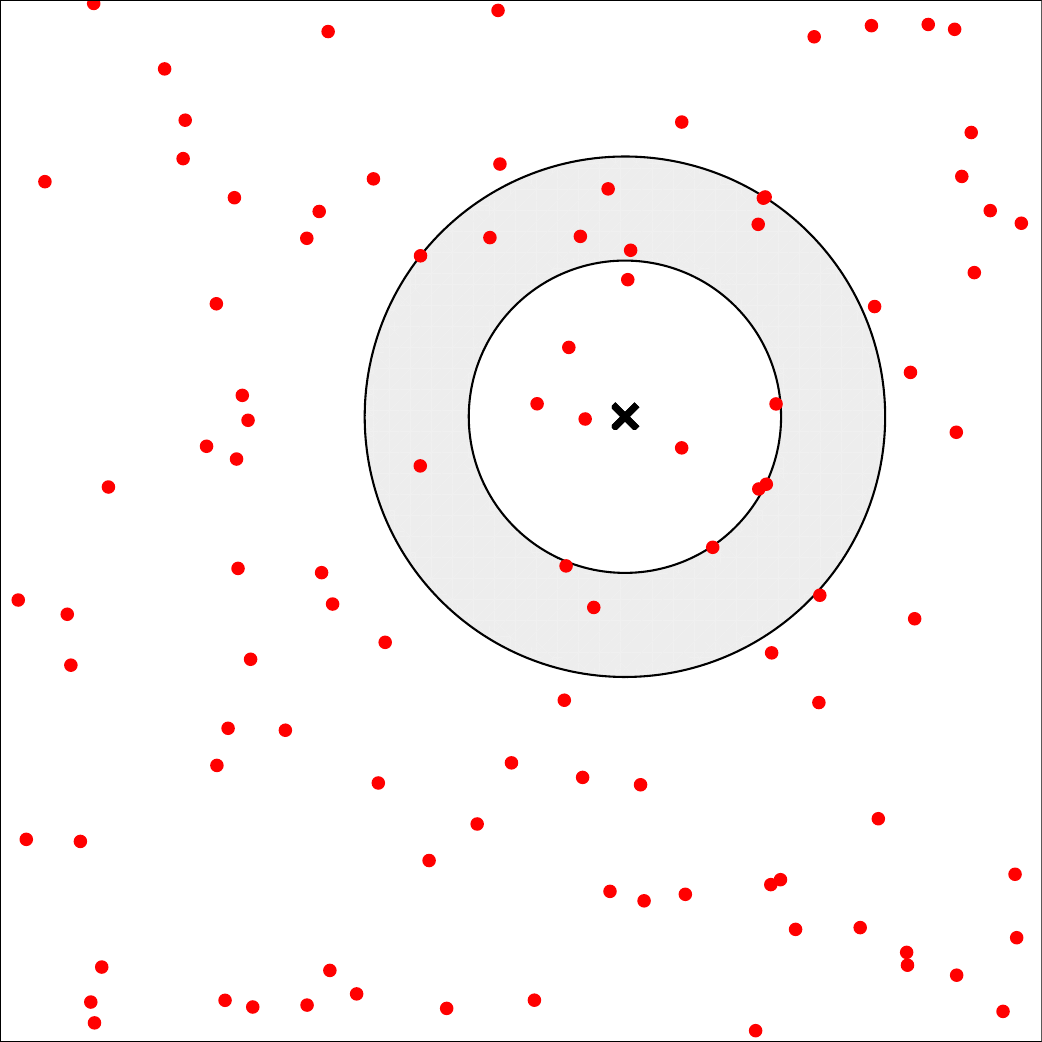}}
\caption{For a galaxy-galaxy correlation function we look in rings of a certain size centered on each galaxy and count the number of galaxies that lie inside each ring. The ring shown is $V(p)$, centered on the galaxy (represented by the black $\times$) having coordinates $p$.}
\label{Vdefgal}
\end{figure}

	We will employ the 2-point function in a similar way to its use in galaxy-galaxy correlation studies. The galaxies correspond to what we have called events. To calculate the galaxy 2-point function for a particular angular separation $\theta$ one counts the number of pairs of galaxies in the sky map whose angular separation is between $\theta$ and $\theta + \Delta \theta$. That is, for every galaxy one looks in a ring of radius $\theta$ and width $\Delta \theta$ around the galaxy and counts the number of other galaxies in this ring. The count is denoted by $C(\theta, \theta + \Delta \theta)(p)$, where $p$ is an index labeling the central galaxy (see Fig.~\ref{Vdefgal}). If the events  were distributed randomly one expects to find on average $\r V(\theta,\theta + \Delta \theta)$ galaxies in this ring, where $\r$ is the overall density of galaxies (number of galaxies in the sky map divided by the area of the map) and $V(\theta,\theta + \Delta \theta)$ is the area of the ring, equal to $2 \pi ({\rm cos}(\theta) - {\rm cos}(\theta +\Delta \theta))$. One then computes the correlation function $\xi$ at separation $\theta$ according to
\begin{equation}
\xi(\theta, \theta+\Delta \theta) = \left \langle \frac{C(\theta, \theta + \Delta \theta)(p) - \r V(\theta,\theta + \Delta \theta)}{\r V(\theta,\theta + \Delta \theta)} \right \rangle,
\label{Xdefgal}
\end{equation}
where the average is taken over the index $p$ of each galaxy. The correlation function $\xi(\theta, \theta+\Delta \theta)$ is interpreted as the fractional increase in probability (above random) that there is a galaxy in a ring between $\theta$ and $\theta +\Delta \theta$ around any given galaxy. This is most easily seen by rearranging (\ref{Xdefgal}) into the form $C = \r V (1 +\xi)$. Notice that the correlation function is inherently a function of the shape and size of the ring in which the search for pairs of events is performed. 

\begin{figure*}
\centering
\resizebox{2.7in}{!}
{\includegraphics{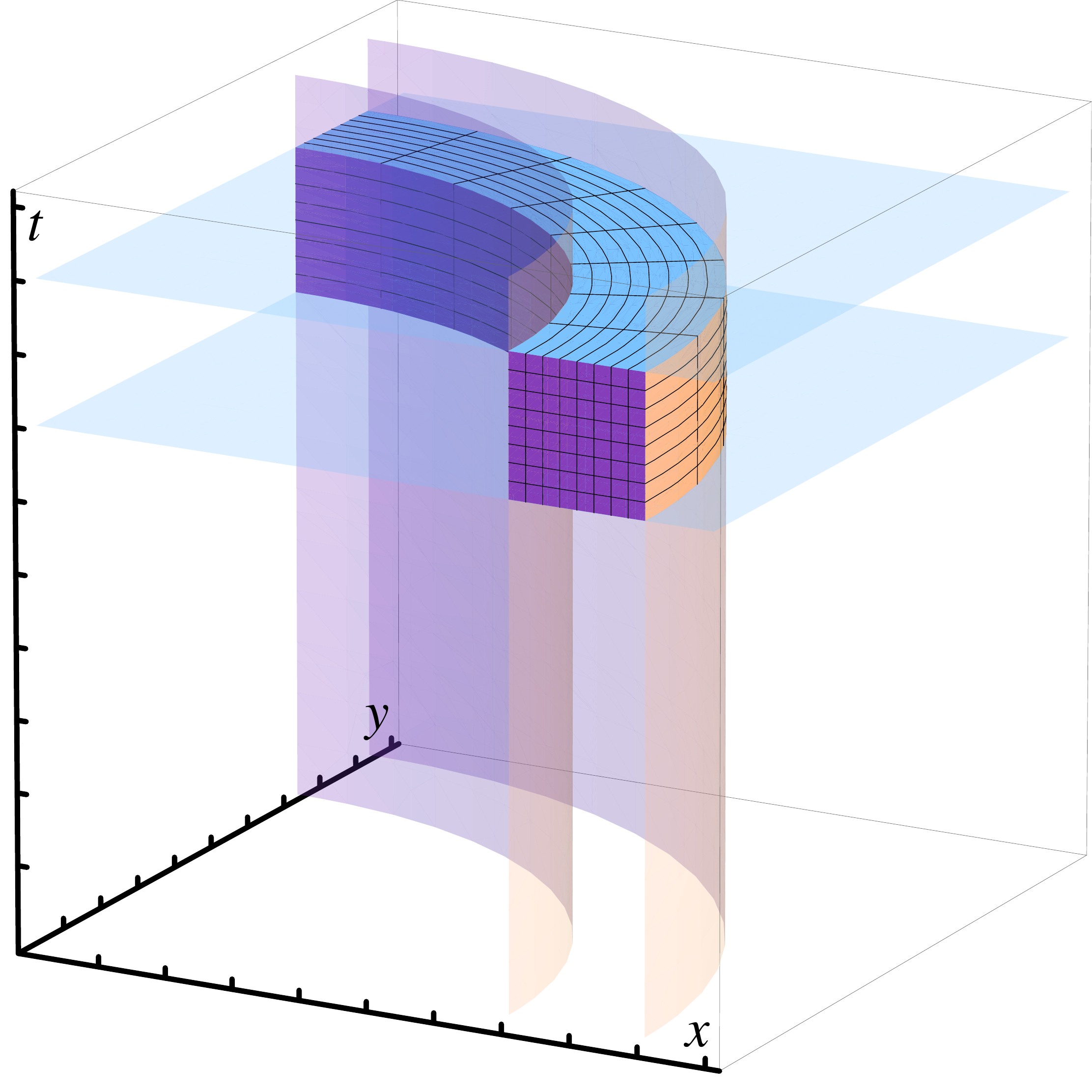}} \quad \quad \quad \quad
\resizebox{2.7in}{!}
{\includegraphics{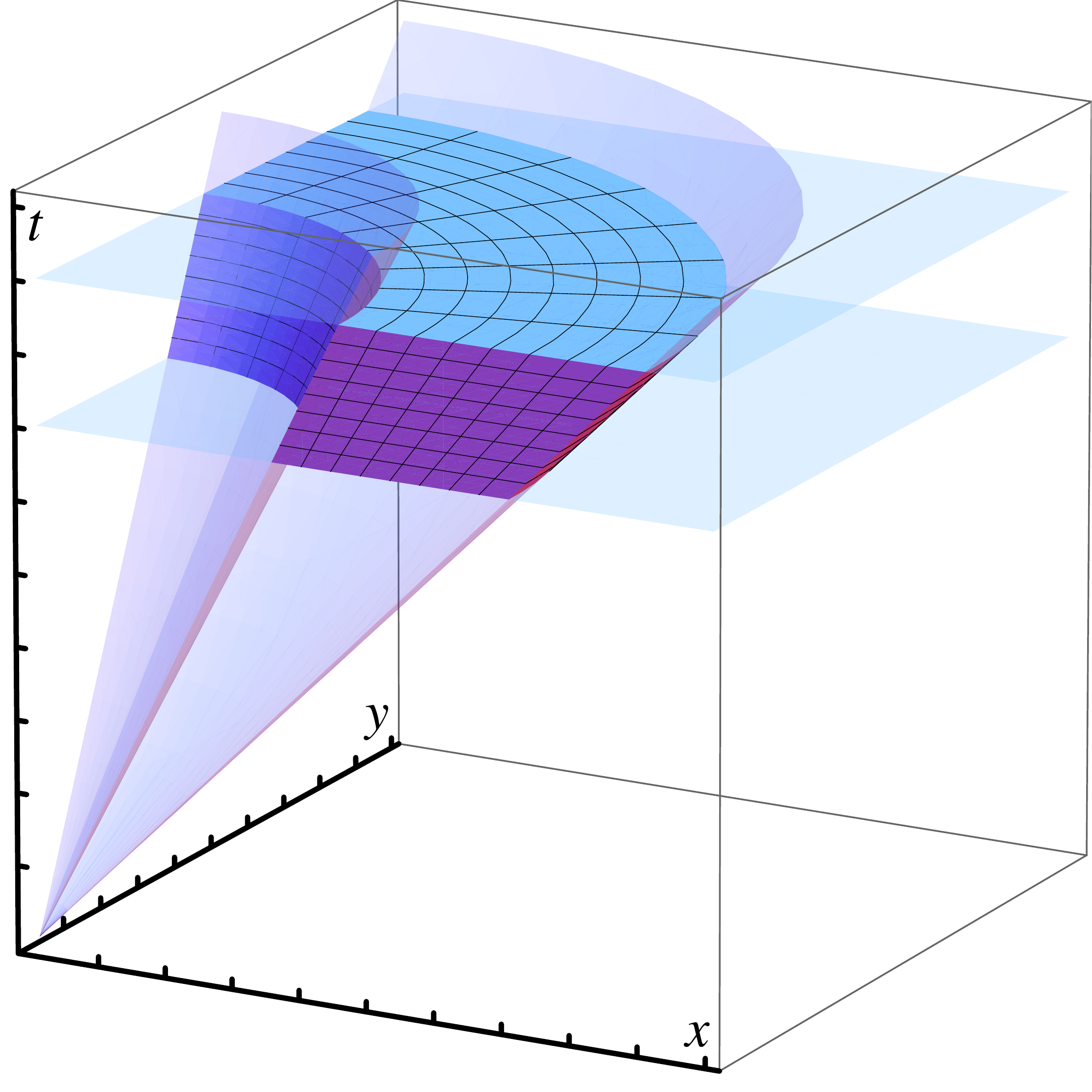}}
\caption{The spacetime regions $V(r_1,r_2;\,t_1,t_2)(p)$ and $\Vvt(p)$ where $p$ is at the origin. The vertical axis is time and the horizontal axes are the $x$ and $y$ coordinates. In the left-hand figure the region between the two cylinders contains all events which have a radial distance from $p$ between $r_1$ and $r_2$. In the right-hand figure the region between the two cones represents the possible worldlines of an object starting at $p$ and having a speed between $v_1$ and $v_2$. Imposing a time separation between $t_1$ and $t_2$ gives the filled regions. }
\label{fig:Vdef}
\end{figure*}

	Now we apply the 2-point function in our situation. We denote spacetime by ${\cal S}$ and we label spacetime events with the abstract index $p$, which carries all the information we have about the event. For example, for the event $p$, $p(t)$ is the time the event occurred, $p(x)$ is the $x$-coordinate of the event, etc. We define the spacetime 2-point function as follows.  For an event at $p$, let $V(p) \subset \cal S$ denote some volume of spacetime which is analogous to the shaded region in Fig.~\ref{Vdefgal}. When there is no confusion $V(p)$ may also refer to the spacetime volume of the region $V(p)$. Two choices for $V(p)$ are illustrated in Fig.~\ref{fig:Vdef}. Let the number of events that occur within the region $V(p)$ be denoted by $C(p)$. When it is important to remember that $C(p)$ depends on the region $V(p)$ we will write it as $C(p;V)$. The spacetime 2-point function is then given by
\begin{equation}
\xi(V) \equiv \left \langle \frac{C(p;V) - \r V(p)}{\r V(p)} \right \rangle,
\label{Xgeneral}
\end{equation}
where the average\footnote{In order to be thorough we should really define $\xi$ by $\xi(p;V) \equiv  \langle [ C(p;V) - \r V(p) ] / \r V(p)  \rangle_{\rm U}$, where the average is taken over an ensemble of Universes. Then we assume that our physical situation is spacetime translation invariant so that $\xi(p;V)$ actually does not depend on the location $p$. Finally, in order to {\em estimate} $\xi$ from a set of data we claim that the average of $\xi(p;V)$ over an ensemble of Universes is equal to the average taken over all the events in our dataset. These are exactly the assumptions which must be made in the theory of galaxy $n$-point functions (referred to as ergodic conditions). We will have more to say on the subject of estimators below.} is taken over every event in the sky map (i.e. over $p$) and $V(p)$ denotes the spacetime volume of the region $V(p)$. As before, $\r$ is equal to the overall spacetime density of events (total number of events divided by the spacetime volume of the sky map). In a realistic application $\r$ will have dimensions of flux: events per square degree per time. 

If the events were all generated by a completely random Poisson process we would expect $C(p;V) = \r V(p)$ on average and $\xi(V)$ would be $0$. The 2-point function $\xi(V)$ is therefore to be interpreted as the fractional probability above random that the region $V(p)$ contains an event given that there is an event at $p$. In the rest of this paper we will develop a formalism for deriving $\xi(V)$.

\section{2-dimensional model}
\label{sec:2d}

\subsection{Ingredients}

	Consider objects moving over a two dimensional surface with constant speeds and each having the same event rate (a ``blinking rate'', so to speak). Each event is then associated with an $x$, $y$, and $t$ value and our ``sky map'' consists of the list of $(x,y,t)$ for each event. The ``blinking'' of an object is a Poisson process with mean rate $\l$: during the time $dt$ each object has a $\l dt$ chance of generating an event. Let the average density of objects be given by $n$, which has units of objects per area. The objects have speeds drawn from the distribution $\pv(v)$: the probability for any given object to have a speed between $v$ and $v + dv$ is $\pv(v)dv$. We consider the case where the velocity distribution is isotropic (accommodating the more general case, $\pvv(\vec{v})d^2\vec{v}$, is straightforward). Finally, some fraction of the events will come from a random (Poisson) component with spacetime density $\ro$: there is a $\ro \,dx \, dy\,dt$ probability of having such an event in any spacetime volume $dx\,dy\,dt$. 

\subsection{The form of $V(p)$ in 2 dimensions}

	There are many possible choices for the spacetime region $V(p)$. The simplest one is
\begin{eqnarray}
V(r_1,r_2;\,t_1,t_2)(p) &\equiv& \{ p' \in {\cal S} :
t_1 \leq p'(t)-p(t) < t_2 \nonumber \\ 
 && \wedge \, \, r_1 \leq d(p',p) < r_2 \},
\label{Vdefrect}
\end{eqnarray}
where $d(p',p)$ is the spatial separation of spacetime events $p$ and $p'$ and $\wedge$ is the logical AND operator. This volume corresponds to all the events whose temporal separation from $p$ is between $t_1$ and $t_2$ and whose spatial separation is between $r_1$ and $r_2$: a ring in spacetime with rectangular cross section (see left panel of Fig.~\ref{fig:Vdef}). The volume of such a region is simply
\begin{equation}
V(r_1,r_2;\,t_1,t_2)=\pi(r_2^2 - r_1^2)(t_2-t_1).
\label{V2drect}
\end{equation}
A more convenient choice for $V(p)$ is
\begin{eqnarray}
\Vvt(p) &\equiv& \{ p' \in {\cal S} : 
t_1  \leq p'(t)-p(t) < t_2 \nonumber  \\
&&  \wedge \, \,  v_1 \leq \frac{d(p',p)}{\left| p'(t)-p(t) \right|} < v_2 \}.
\label{Vdef}
\end{eqnarray}
This region is interpreted as the volume of spacetime that an object might explore between time $t_1$ and $t_2$ if it started at $p$ and had any speed in the range from $v_1$ to $v_2$. $\Vvt(p)$ is illustrated in the right panel of Fig.~\ref{fig:Vdef} for the case where $p$ is at the origin of spacetime coordinates. If the event was an object which had velocity between $v_1$ and $v_2$ its worldline would lie between the cones $x^2 +y^2 = v_1 t^2$ and $x^2 +y^2 = v_2 t^2$ and we only consider the region where $t_1 \leq \left| p'(t)-p(t) \right| <t_2$. The volume of this region may be found by slicing the shaded region in the $x-t$ plane, and rotating each small piece around the $t-$axis. The result is
\begin{equation}
\Vvt = \int\limits_{t_1}^{t_2} \int\limits_{v_1 t}^{v_2 t} 2 \pi x dx dt = \frac{\pi}{3}(v_2^2 -v_1^2)(t_2^3 - t_1^3).
\label{V2d}
\end{equation}

\subsection{Coordinate systems}

The forms of $V(p)$ presented in the previous section are easiest to visualize in the cartesian coordinate system $(x,y,t)$ (see Fig~\ref{fig:Vdef}). However, a more appropriate choice of spacetime coordinates are $(v,t,\phi)$, defined by
\begin{eqnarray}
v    &=&\sqrt{(x^2 + y^2)}/ |t| \nonumber \\
t    &=&t \nonumber \\
\phi &=& {\rm \tan^{-1}}(y/x).
\label{vtphicoords}
\end{eqnarray}
These are just cylindrical coordinates $(r, z, \phi)$ but with $r$ scaled by the absolute value of $z$. The Jacobian for this change of variables is
\begin{equation}
dV(v,t,\phi) \equiv dx\,dy\,dt = vt^2 dv\,dt\,d\phi.
\label{dV2d}
\end{equation}
The volume $V$ of any region of spacetime $V(p)$ is given by
\begin{equation}
V = \int\limits_{V(p)} dV(v,t,\phi) = \int\limits_{V(p)}vt^2 dv\,dt\,d\phi \,.
\label{Vgeneral}
\end{equation}
For example, we can recover (\ref{V2drect}) as 
\begin{displaymath} 
\nonumber 
V(r_1,r_2;\,t_1,t_2) = \int_{t_1}^{t_2}\int_{r_1/t}^{r_2/t}\int_{0}^{2\pi} dV(v,t,\phi)
\end{displaymath} 
and (~\ref{V2d}) as 
\begin{displaymath}
\nonumber
\Vvt = \int_{t_1}^{t_2}\int_{v_1}^{v_2}\int_{0}^{2\pi} dV(v,t,\phi).  
\end{displaymath} 

For later use we define the $(v,t,\phi)_p$ coordinate system which is the same as the coordinate system described in (\ref{vtphicoords}) except the center of coordinates $(v=0,\, t\rightarrow 0)$ is at the spacetime point $p$. We also define the corresponding volume element $dV_p(v,t,\phi)$ or $dV_p$ for short. The region $dV_p(v,t,\phi)$ is the infinitesimal version of (\ref{Vdef}), i.e $dV_p(v,t,\phi) = V(v, v+dv;\, t,t+dt)$, only not rotated about the t-axis.

	Finally we should note that it is just as easy to derive our results for rectangular coordinates. One just uses the coordinate system $(v_x, v_y, t)$ where
\begin{eqnarray}
v_x &=& x/t \nonumber \\
v_y &=& y/t \nonumber \\
t   &=& t,
\label{vxvytcoords}
\end{eqnarray}
instead of (\ref{vtphicoords}). The analogue of (\ref{dV2d}) is then 
\begin{equation}
dV(v_x,v_y,t) \equiv dx\,dy\,dt =  t^2 dv_x\,dv_y\,dt.
\label{dV2drect}
\end{equation}

\subsection{The form of $\xi(V)$ in 2 dimensions}
Our goal is to write down an analytic expression for (\ref{Xgeneral}). In the previous subsection we showed  how to calculate the volume $V(p)$. Now we move on to $C(p;V)$, which can be interpreted as follows. Given an event at $p$, $C(p;V)$ is the probability\footnote{Or, if it is greater than $1$, $C(p;V)$ is the expected number of events in the region $V(p)$.} of finding another event in the spacetime region $V(p)$.

There are two processes by which an event might occur in $V(p)$. Accordingly, we can break up $C(p;V)$ into the sum of two terms: $C(p;V) = $ [the probability of getting an event from an object that was at p] + [the probability of getting an event from any other source]. The first term can be thought about in a series of steps: given an event at $p$ find the probability that it came from an object, that this object moves into the region $V(p)$, and that this object triggers a new event while in this region.

The probability $\p1$ that any given event came from a moving object (as opposed to being generated by the Poisson component of the background) is the ratio of the flux from moving objects to the total flux:
\begin{equation}
\p1 = \frac{ n \l}{n\l+\ro} = \frac{ \r_1}{\r},
\label{p1def}
\end{equation}
where $\r_1 \equiv n \l$ is the average flux of the moving objects. As before, $\r$ is the total spacetime density of all events (i.e. the overall flux).

	The probability that the object moves into the region $dV_p(v,t,\phi)$ is simply the probability that its speed is between $v$ and $v+dv$, $\pv(v)dv$, multiplied by $d\phi /2\pi$, the probability that it is moving in a direction between $\phi$ and $\phi + d\phi$ \footnote{If the objects do not have an isotropic velocity distribution then this probability is ${\rm P}_{\vec{v}}(v,\phi) dv d\phi$, where ${\rm P}_{\vec{v}}(v,\phi)$ is the probability density for the velocity {\em vector}.}. If the object makes it into the region $V(p)$ the probability of it generating a second event is $\l dt$. Therefore, the probability that there was an object at $p$ which moved into $V(p)$ and generated another event is\footnote{In full generality this equation would be $\p1 \int_{V(p)} {\rm P}_{\vec{v}}(v,\phi)\l(t)\, dv \, dt \, d\phi$, where $\l(t)dt$ is the probability that an object which generated an event at $t=0$ generates another event in the time interval between $t$ and $t+dt$.}. 
\begin{equation}
\p1 \int\limits_{V(p)} \pv(v)\, dv\, \l dt\, \frac{d\phi}{2\pi}. 
\label{eq:p1}
\end{equation}
	The second term in $C(p;V)$ is simply $\r V(p)$, where $\r \,dx \, dy\,dt$ is the probability that any random spacetime volume $dx \, dy\,dt$ contains an event from either an object or the random component (note that $\r = \r_1 + \ro$). 
	
Therefore, putting together these parts and plugging them into (\ref{Xgeneral}) we find
\begin{eqnarray}
\xi(V) &=& \left \langle \frac{C(p;V) - \r V(p)}{\r V(p)} \right \rangle \nonumber \\
       &=& \frac{\p1 \int_{V(p)} \pv(v)\, dv\, \l dt\, d\phi/ 2\pi}{\r V(p)} \label{XV2dgen}\nonumber \\ 
       &=& \frac{\r_1 \, \int_{V(p)} \pv(v)\, dv\, \l dt\, d\phi / 2\pi}{(\r_1 + \ro)^2 \, V(p)}.
\label{XV2d}
\end{eqnarray}

As is usually done for galaxy-galaxy correlation functions let's see what happens when we take the limit $V(p) \rightarrow dV_p(v,t,\phi)$. Using (\ref{dV2d}) we see that
\begin{equation}
\xi(V) \rightarrow \xi[dV(v,t,\phi)] = \frac{\r_1 \l}{2 \pi (\r_1 + \ro)^2} \frac{ \pv(v)}{vt^2}.
\label{XdV2d}
\end{equation}

	This limit is finite and the function $\xi$ traces the velocity distribution of the population of objects. Therefore, if $\xi$ can be measured for multiple values of $v$ then it is possible to directly reconstruct both the velocity distribution of the moving objects and information about their abundance and luminosity.

\section{The error in $\xi$}
\label{sec:errors}

	Getting a handle on the error $\DX$ in a measurement of $\xi(V)$ is just as important as calculating $\xi(V)$ itself: any practical application of this method will reveal nothing if the uncertainty in $\xi(V)$ is comparable to $\xi(V)$. The zeroth order discovery that can be made using the 2-point function is the detection of the presence of moving objects. This is done by rejecting the hypothesis that $\xi(V) = 0$ for all choices of $V(p)$, which is possible only if $\DX / \xi(V) < 1$ for some choices of $V(p)$. An estimate of the error is also essential when fitting the theoretical value for $\xi$ to the data; i.e. when performing a $\chi^2$ fit to determine the physical parameters describing the density, luminosity, and velocity distribution of contributing sources. 

	Fortunately, the errors in correlation functions have been thoroughly studied in the case of galaxy-galaxy correlations \citep{1973ApJ...185..413P,1977ApJ...216..682F,1992ApJ...392..452M,Fry:1992vr,1993ApJ...412...64L,1993ApJ...417...19H,1994ApJ...424..569B,1996ApJ...470..131S,1998ApJ...494L..41S}. We emphasize that all the technology that has been developed for calculating 2-point functions for galaxies and quantifying their errors can (and should) be straightforwardly applied to our 2-point function. As stated before, the only conceptual difference between the two tools is the choice of $V(p)$.

	In particular, we apply the results of \citet{1993ApJ...412...64L} (hereafter LS93) to the present problem. In the examples below we measure $\xi(V)$ using an unbiased estimator which is identical to the $DD/RR$ ratio in LS93. This is done for simplicity. The unbiased LS93 $(DD-2DR+RR)/RR$ estimator was shown to have a smaller variance and should be used in practical applications. (In LS93, $DR$ refers to the cross-correlation of the observed data with sets of completely random data, while $DD$ and $RR$ are the auto-correlation functions computed for the data and for a completely random set of data, respectively.)
	
	To quantify the error in $\xi(V)$ we adapt the LS93 expression for the variance of the $(DD-2DR+RR)/RR$ estimator for small correlations (i.e. small values of $\xi(V)$, likely in cases of physical interest). For a given shape $V(p)$ (with spacetime volume $V$) the variance of the estimator is given by
\begin{equation}
\DX^2(V) = \frac{[1+\xi(V)]^2}{ N \r V},
\label{DX}
\end{equation}
where $N$ is the total number of events in the sky map. This can be seen to be the same as Eqs. 43 and 48 in LS93 by writing $\r = N / {\cal V}$, where ${\cal V}$ is the total spacetime volume of the sky map and noting that $V(p)/{\cal V}$ is equal to LS93's $G_p (\theta)$. The signal to noise ratio is then
\begin{equation}
\frac{\xi(V)}{\DX(V)} = \frac{\xi(V)}{1+\xi(V)} \sqrt{N \r V}.
\label{eq:SNR}
\end{equation}
These expressions should be used to determine the optimal volumes $V(p)$ for any given application. Ideally, $V(p)$ should be chosen to make the signal to noise ratio large while keeping $V(p)$ small enough that many choices for $V(p)$ can be measured for the sky map. This dilemma occurs with galaxy-galaxy correlation studies as well. An annulus of specific size (see Fig.~\ref{Vdefgal}) corresponds to one choice of $V(p)$. One would like to choose the width of the annulus as small as possible so that the correlation function can be measured at many different angular scales. However, the smaller the width of the annulus the larger the uncertainty in the measured value of the 2-point function.

As is well-known in galaxy-galaxy correlation studies, measurements of $\xi$ at two different angular sizes can be highly correlated. This issue will also affect any measurement of the spacetime 2-point function: the measured $\xi(V)$ for different choices of $V(p)$ may be correlated. Therefore, a $\chi^2$ fitting to extract physical parameters should include an estimate of the covariance of $\xi(V)$ between different $V(p)$'s. A variety of methods have been developed to estimate or predict this covariance matrix. Many of these are trivially adapted for use in this case. Bootstrapping (e.g. \cite{1986MNRAS.223P..21L, 1984MNRAS.210P..19B,1994MNRAS.266...50F}) and jackknife resampling \citep{1993stp..book.....L, 2002ApJ...571..172Z} require measuring the correlation function on various subsets of the full data set and analyzing the variation among these estimates of $\xi$. If generating fake data sets is feasible then one can simply measure the correlation function on many fake maps to find the covariance of $\xi(V)$ between various $V(p)$'s.

\section{Point spread function and computational considerations}

	In this section we discuss two ways to include information about the point spread function (PSF) into the derivation of the form of the 2-point function. This will serve as a guide for incorporating the PSF in realistic applications.
	
	The PSF of a detector quantifies the uncertainty in its measurement of the locations of events in spacetime \citep{2004APh....20..485M}. The PSF typically takes the form $\psf$, where $\pt$ is the true location of the event and $\po$ is the location that the detector reports, the ``observed'' location. The PSF is a probability density on spacetime: $\psf dV$ is the probability that if the detector reports an event at $\po$ it actually arrived from the spacetime region $dV$ centered at $\pt$\footnote{Because the time resolution of detectors is generally excellent compared with the spatial (or angular) resolution, the PSF is usually given as a function of spatial coordinates only. The PSF we have defined would then be equal to $\delta(t_{\rm t} - t_{\rm o})\, {\rm PSF}(\vec{r}_{\rm t} - \vec{r}_{\rm o})$.}. As a result, $\int \psf dV_{\rm t} = 1$. Additionally, there is the probability $\pd$ that if a signal (e.g. a photon) arrives at the detector it will actually be detected as an event.
			
	If we are given the PSF for a given event we can do a more precise job of computing $C(p;V)$. As above we want to answer the question: given that the detector reported an event at $\po$ what is the probability that the detector reports another event in the spacetime region $V(p)$? 

	If the detector reports an event at $p$ there is a $\p1$ chance that it received a signal from a moving object. But the true location of the object could be anywhere, with probability given by the PSF. The object can have any velocity and can emit a signal at any later time. This signal has a $\pd$ chance of being detected. The location of the observed event is again determined by the PSF. Specifically, we have
%
\begin{eqnarray}
C(p;V) &=& \p1 \int\limits_{\pt \in {\cal S}} {\rm PSF}(p_{\rm t} - p) \, dV_p (v_{\rm t},t_{\rm t},\phi_{\rm t})  \nonumber \\
&\times&
 \int\limits_{\pt' \in {\cal S}} \frac{1}{2\pi} \pv(v') \l \pd dv' \, dt' \, d\phi' \nonumber \\
 &\times& \int \limits_{\po \in V(p)} {\rm PSF}(\pt' - \po) \, dV_{\pt'}(v_{\rm o},t_{\rm o},\phi_{\rm o}).
\label{Cpsf}
\end{eqnarray}
%
	In words, there is a $\p1$ chance that the observed event at $p$ came from a moving object. Given that it came from a moving object there is a ${\rm PSF}(p_{\rm t} - p) \, dV_p (v_{\rm t},t_{\rm t},\phi_{\rm t})$ chance the event actually occurred in the region $dV_p (v_{\rm t},t_{\rm t},\phi_{\rm t})$ around the point $\pt = (v_{\rm t},t_{\rm t},\phi_{\rm t})_p$ (recall the definition of $dV_p$ at the end of the section on the choice of $V(p)$).
Then there is a $(1/2\pi) \pv(v') \l \pd \, dv' \, dt' \, d\phi'$ chance that the object moves into the region $dV_{\pt}(v',t',\phi')$ around the point $\pt' = (v',t',\phi')_{\pt}$ and emits a signal which is reported by the detector. Finally there is a ${\rm PSF}(\pt' - \po) \, dV_{\pt'}(v_{\rm o},t_{\rm o},\phi_{\rm o})$ chance that this event is reported as having occured in the region $dV_{\pt'}(v_{\rm o},t_{\rm o},\phi_{\rm o})$ around the point $\po = (v_{\rm o},t_{\rm o},\phi_{\rm o})_{\pt'}$.

	All the possibilities are taken into account by integrating $\pt$ and $\pt'$ over all of spacetime (the object could actually have been located at any point and could have moved to any other point) and by integrating $\po$ over the region $V(p)$ (we are only interested in the possibilities where the detector reports the second event in the region $V(p)$). For clarity we have omitted the $\r V(p)$ term in $C(p;V)$, which represents the probability of a reported event in $V(p)$ from any source besides an object moving from $p$ into $V(p)$. One can show that (\ref{Cpsf}) reduces to the numerator of (\ref{XV2dgen}) when $\psf = \delta(\pt - \po)$ and $\pd=1$.

The spacetime correlation function is an example of a 2-point correlation function and so any method that is used to compute 2-point functions may also be used here. In galaxy-galaxy studies, the galaxies are localized sources and the 2-point function is measured by counting pairs of galaxies which have a particular separation. When looking for moving objects using gamma-ray data, for instance, the events are also localized. Computational procedures then carry over directly. Typically, counting pairs of events is an $N^2$ process\footnote{We point out that efficient algorithms with better than $N^2$ scaling have recently been developed. See, for example, \cite{2001misk.conf...71M,2005NewA...10..569Z,2004ApJS..151....1E}.}. For example, in gamma-ray diffuse studies the number of events is proportional to the observation time as well as to the effective area of the detector.

In other situations the data do not come as localized events but as a continuous amplitude across the sky. This case can be treated by first discretizing the survey area into small ``cells'' or pixels. Each pixel now has a continuous value. For a particular $V(p)$, the correlation function is found by multiplying the value of the pixel $p$ by the sum of the values in the pixels in the volume $V(p)$. The expected value of this quantity (i.e. the denominator in (\ref{Xgeneral})) is the average pixel amplitude squared multiplied by the volume of $V(p)$.

This method of computing the 2-point function can be used as an alternative way to account for the detector point spread function. Following \citet{2004APh....20..485M}, every discrete photon event in spacetime is replaced by its point spread function. The overlap of the point spread functions for all observed events forms a continuous density over the survey area and observation window. The 2-point correlation function for any choice of $V(p)$ can then be measured as described above. We note that this method suffers no performance penalty for increased numbers of observed events because the events are essentially binned into pixels in spacetime, with each pixel having a value given by the linear superposition of all contributing PSFs.

\section{Examples of the 2-Dimensional formalism}
\label{sec:demos}

In this section we will demonstrate the accuracy of the derivations by measuring $\xi$ for three different simulations in which the objects move according to a specific speed distribution. A generic choice for $\pv$ is the Rayleigh distribution: the speed distribution for a 2-dimensional isotropic Gaussian velocity distribution. It has the form
\begin{equation}
\pv(v)=\frac{v}{a^2} \, e^{-v^2 / 2a^2},
\end{equation}
with mean speed $\bar{v} = a\sqrt{\pi/2}$.  We choose $V(p)$ to be the region described by (\ref{Vdef}) and shown in the right panel of Fig.~\ref{fig:Vdef}. We note that any choice of the shape of $V(p)$ is allowed. 
The shape $V(p)$ used here is adapted to the search for objects which move in straight lines at constant speed. For sources with different patterns of motion, other choices for $V(p)$ may be more appropriate. However,  because the choice affects the counting of pairs of events when measuring $\xi$ it must be taken into account in the theoretical derivation of $\xi$.

With these choices the integral in (\ref{XV2d}) becomes
\begin{eqnarray*}
\int_{V(p)} \pv(v)\, dv\, dt\, \frac{d\phi}{2\pi} &=& \int\limits_{t_1}^{t_2}\int\limits_{v_1}^{v_2}\int\limits_{0}^{2\pi} \frac{v}{a^2} e^{-v^2 / 2a^2} dv\, dt\, \frac{d\phi}{2\pi} \nonumber \\
                                                  &=& (t_2-t_1) \left[ e^{-v_1^2 / 2a^2} - e^{-v_2^2 / 2a^2} \right].
\end{eqnarray*}
Inserting this expression into (\ref{XV2d}) and using (\ref{V2d}) we find
\begin{equation}
\Xvt= \frac{\r_1 \l \, (t_2-t_1) \left[ e^{-v_1^2 / 2a^2} - e^{-v_2^2 / 2a^2} \right]}{(\pi/3) (\r_1 + \ro)^2 \, (v_2^2 -v_1^2)(t_2^3 - t_1^3)}.
\label{Xvt2d}
\end{equation}

Given an event map we can measure $\Xvt$ for any choice of the 4 parameters $(v_1, v_2, t_1, t_2)$. In practice, a fit can be attempted in order to discover the 4 physical parameters $\l,\, \r_1,\, \ro$, and $a$. While $\r = \r_1 +\ro$ is measured directly the parameters $\r_1$ and $\l$ are combined as a single normalization factor and so the most a fitting analysis would reveal would be the combination $\r_1 \l$. In the 2-dimensional case this is true for any choice of $V(p)$, as can be seen from (\ref{XV2d}). Of course, knowledge of any one of $\l, \r_1,$ or $\ro$ can be used to find the other two.

\begin{figure}
\centering
\resizebox{3.3in}{!}
{\includegraphics{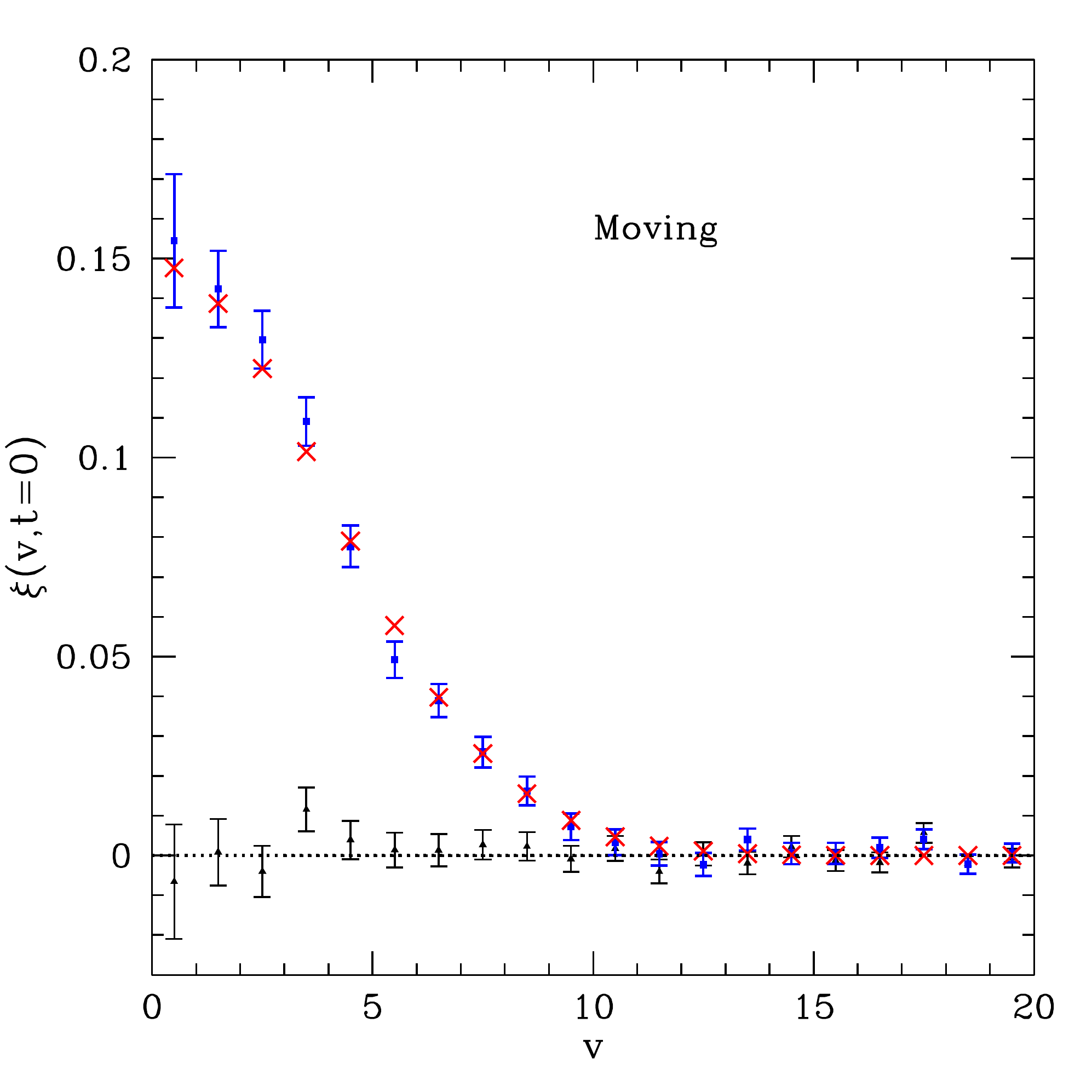}}
\caption{A toy example demonstrating the use of the spacetime correlation function to discover the presence of localized event sources with non-zero speeds. The $t=0$ slice of $\xi(v,t)$ is plotted showing the theoretical prediction (red $\times$'s), the measured value (blue squares), and the measured value for the case of completely random events (black triangles). The hypothesis that the pattern of events in the sky map is Poisson ($\xi(v,t) = 0$) is clearly rejected at high significance. The error bars in the measured quantities are explained in the discussion surrounding (\ref{DX}). The sky map contained 3.5 million events, all from moving objects, though each object contributed only 0.1 events on average. The blue data points are measured from a larger version of the map shown in the central panel of Fig.~\ref{fig:2limits} while the black points are measured from a larger version of the map shown in the right panel.}
\label{fig:xi_M}
\end{figure}

	Our first simulation will only contain moving objects. In the second we will add a component of random noise and in the third simulation both random noise and a population of stationary objects will be considered in addition to the moving objects. One of the goals of these simulations is to demonstrate that the 2-point function can tell the difference between a background containing a population of sources and a completely random background. We do this by generating a second sky map for each example with the same number of events but distributed completely randomly throughout the spacetime volume. The 2-point function is measured for this randomly generated sky map and is plotted along with the 2-point function measured from the actual simulation. If the events are randomly generated there should be no correlations at all: $\Xvt$ should be $0$ for all values of $v_1, v_2, t_1$, and $t_2$. 

\subsection{Example 1: Moving sources only}

	In the first simulation there is no random noise: $\r_0 = 0$. We simulate an area with dimensions 13,200 $\times$ 13,200 for time 10. The density of objects is $n = 0.2$ and each has an event rate $\l = 0.01$ yielding an estimated flux of $\r_1 = n \l = 0.002$ events per unit area per unit time. Their speeds are distributed according to a Rayleigh distribution with a mean speed $\bar{v} = 5$. The objects then have the same density, event rate, and speeds as in the middle image of Fig.~\ref{fig:2limits}. The expected number of events triggered by each object is $0.1$ which means that although there are about 35 million objects present, less than $10\%$ of them will trigger even a single event. Overall, there are roughly 3.5 million events in our sky map.
	
		In the measurement of $\Xvt$ we take $v_2 = v_1 + 1$ and $t_2 = t_1 + 1$, i.e. we choose non-overlapping bins of size 1 in both time and velocity separation. The subscripts on $v_1$ and $t_1$ are dropped and $\Xvt$ is relabeled $\xi(v,t)$. The 2-point function is then measured for $v = 0, \dots, 19$ and for $t = 0, 1, 2$. The $t=0$ slice of the measured $\xi(v,t)$ is shown in blue in Fig.~\ref{fig:xi_M} along with the theoretical value (\ref{Xvt2d}), shown in red. The black curve is the 2-point function measured for a sky map containing the same number of events but placed randomly. The separation $v_2-v_1 = 1$ was selected for illustrative purposes. The time separation $t_2-t_1 = 1$ was then chosen to be close to the optimal separation found by maximizing the signal to noise ratio (\ref{eq:SNR}) for $t_1=0$. The error bars are computed according to (\ref{DX}). This is a slight abuse since the estimator plotted is $DD/RR$ and not $(DD-2DR+RR)/RR$. In practice it is recommended to use the latter estimator.		
		
		It is clear that moving objects are detected at a very high significance (i.e. the hypothesis $\xi(v,t)=0$ is rejected). The measured value $\xi(0,0)=0.15$, for example, is about 15 standard deviations from $\xi=0$. A fit to recover the parameters $\l, \r_1$, and $a$ can be attempted using $\xi(v,t)$, which is measured at the lattice of points $\{(v,t): v=0,1,\dots ; t=0,1,\dots \}$. In practice, the full covariance matrix of errors between different $v$-bins should be included in such a fit (see last paragraph in Sec.~\ref{sec:errors}).

\subsection{Example 2: Moving sources and a random component}	
	
	Let us see if the spacetime 2-point function can tell the difference between a collection of moving objects plus random noise and a situation with just random noise, where both cases have the same total flux $\r$. The sky map has the same dimensions as before and the moving objects have the same number density, luminosity, and speed distribution as before yielding $\r_1 = 0.002$. We choose the random component to have the same flux $\ro=\r_1$ so that $\r = \r_0 + \r_1 = 0.004$. There are about 7 million events in this sky map, half coming from objects and the other half coming from the random component.

\begin{figure*}
\centering
\resizebox{3.3in}{!}
{\includegraphics{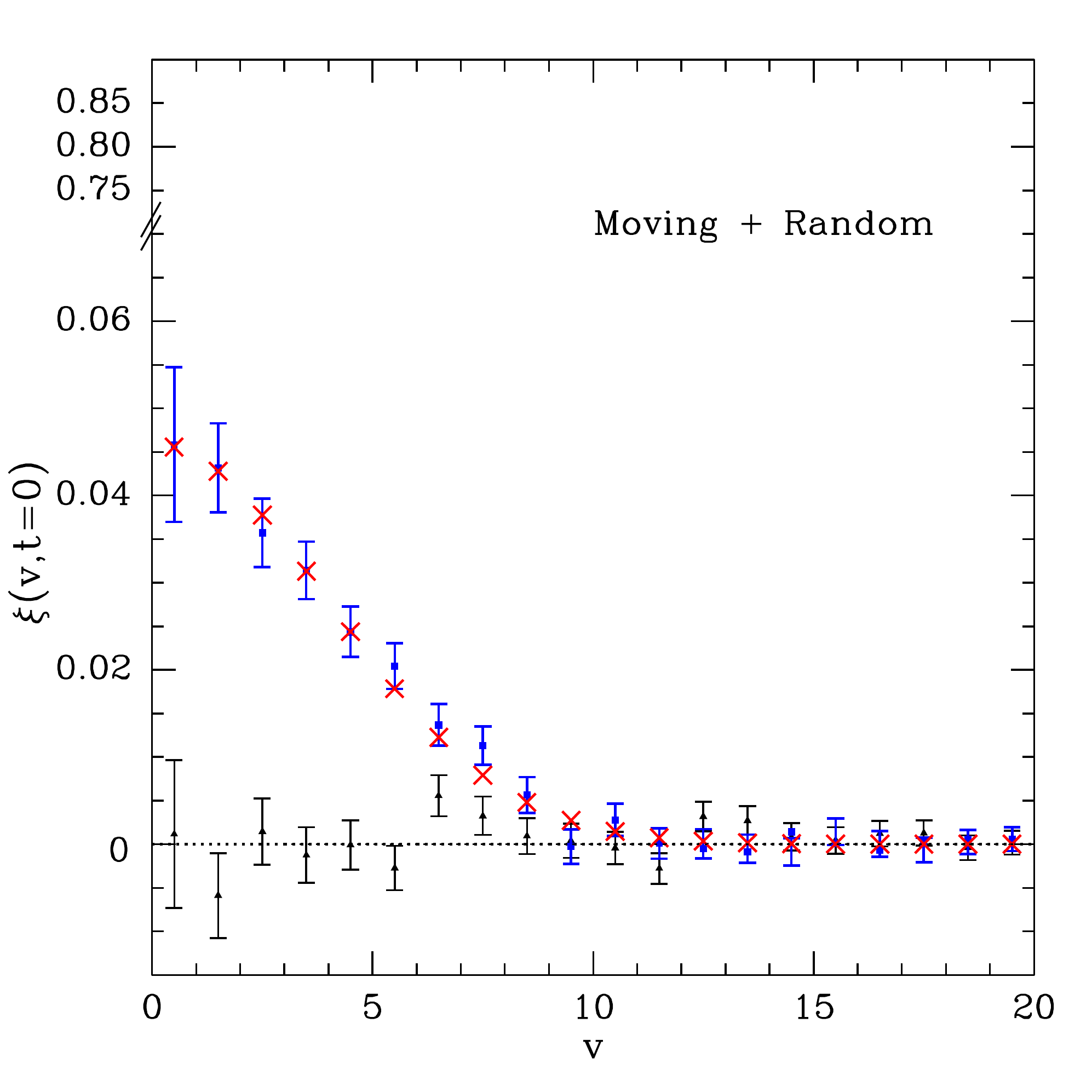}}
\resizebox{3.3in}{!}
{\includegraphics{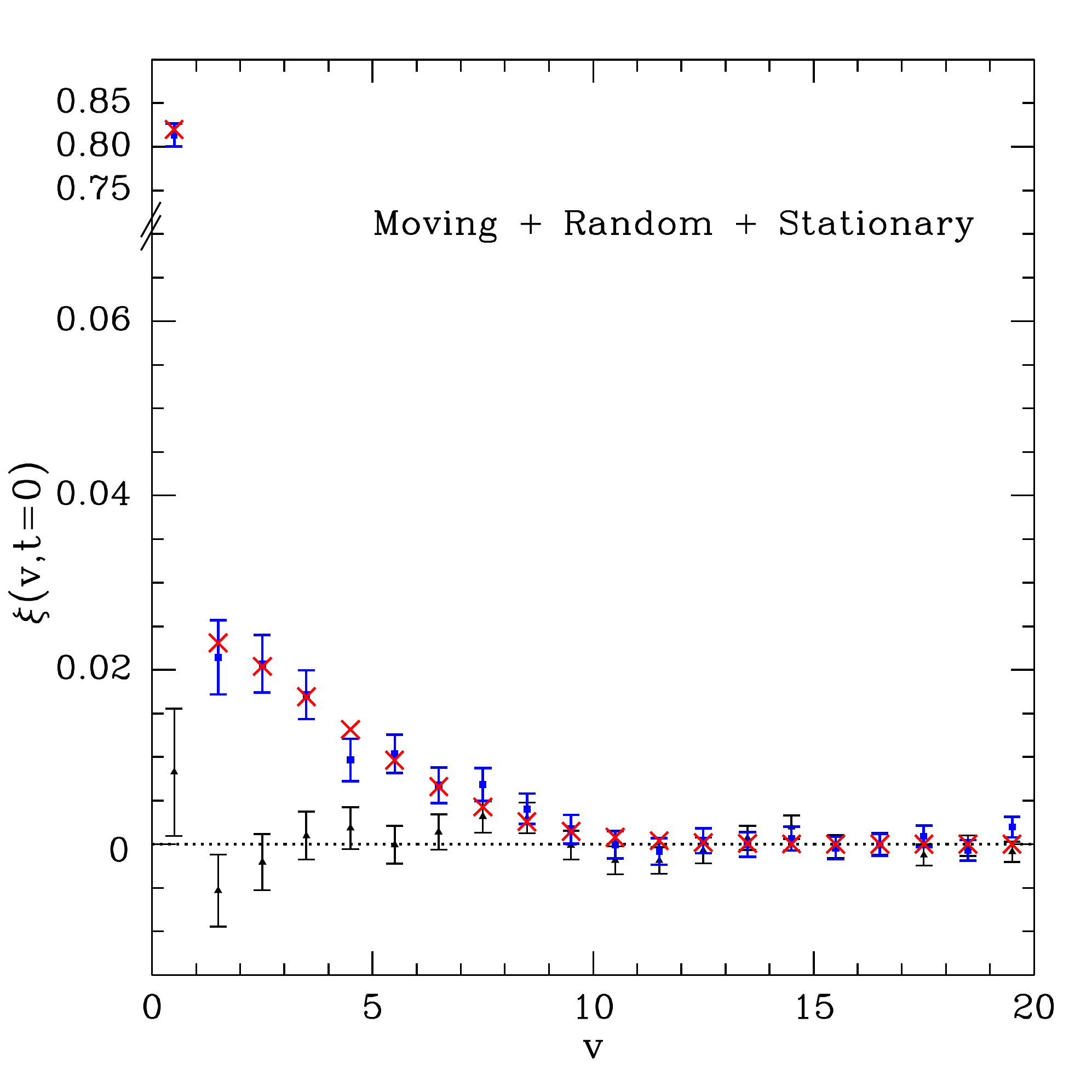}}
\caption{The 2-point function $\xi$ measured for two simulations color-coded as in Fig.~\ref{fig:xi_M}. Each contained 7 million events. Objects had the same event rate as the first simulation. {\em Left:} Moving sources and random noise. Half the events came from moving objects and half were generated completely randomly to represent noise.
{\em Right:} Moving sources, stationary sources and random noise. A third of the events are from moving objects, a third from stationary objects, and the last third were generated randomly.}
\label{fig:xi_MR_MRS}
\end{figure*}
		
	In the calculation of $\Xvt$ we choose $v_2 = v_1 + 1$ and $t_2 = t_1 + 0.9$. 
	The results are plotted in the left panel of Fig.~\ref{fig:xi_MR_MRS}. Again it is clear that the moving objects are detected even in the presence of random signal in the sky map.
		
	How impressive is this result? Could we have just looked at the data by eye and spotted the presence of moving objects? If each object generates at most a single event then clearly it is impossible to determine anything about their motion or to distinguish this from the case of completely random events. The fraction of events which come from objects that trigger more than one event is
\begin{eqnarray}
{\rm P}_{>1} &=& \frac{nA \sum\limits_{k=2}^\infty k \,\, \pi(k;\l T)}{(\r_1 +\ro)AT} \nonumber \\
       &=& \frac{nA \,\,\, \l T (1- e^{-\l T})}{(\r_1 +\ro)AT} \nonumber \\
       &=& \frac{\r_1}{\r_1+\ro}(1- e^{-\l T}),
\label{P>1}
\end{eqnarray}
where $A$ total area of the sky map, $T$ is the observation time, and $\pi(i;M) \equiv e^{-M} M^i / i!$ is the Poisson distribution with mean $M$. In our case, $\r_1 = \ro = 0.002$, $\l = 0.01$, and $T=10$. Substituting these values into (\ref{P>1}) gives ${\rm P}_{>1} =0.048$. That is, less than 5\% of the events in our simulated sky map come from objects which generate more than one event. Furthermore, 95\% of these events come from objects which generate exactly two events during the time $T$. If one was to try to spot individual moving objects in the sky map one would need to be able to take 200 events and out of the nearly 20,000 possible pairs of these events spot the 5 pairs which correspond to an object triggering an event, moving, and triggering a second event.

\begin{figure*}
\centering
\resizebox{3.2in}{!}
{\includegraphics{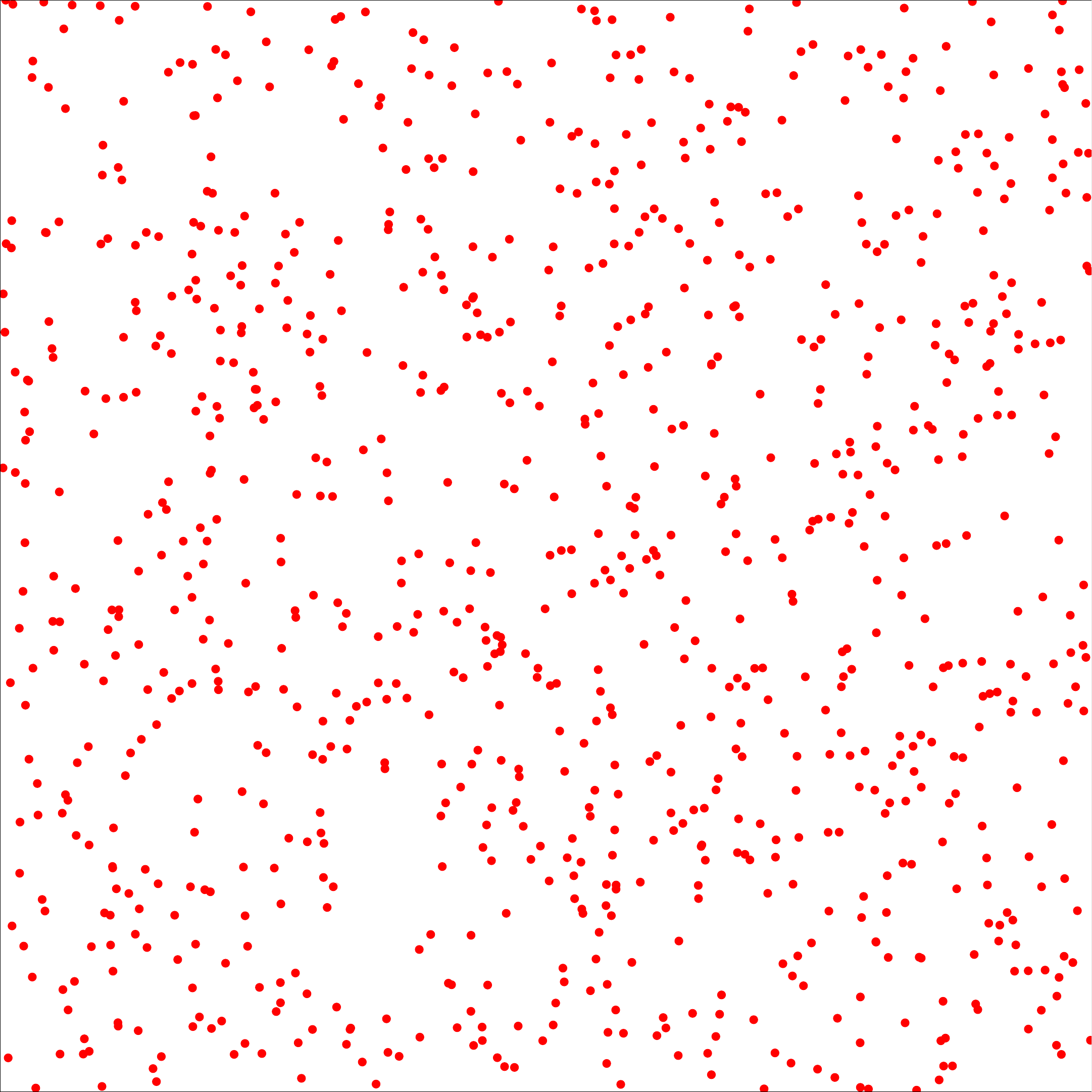}} 
\resizebox{3.2in}{!}
{\includegraphics{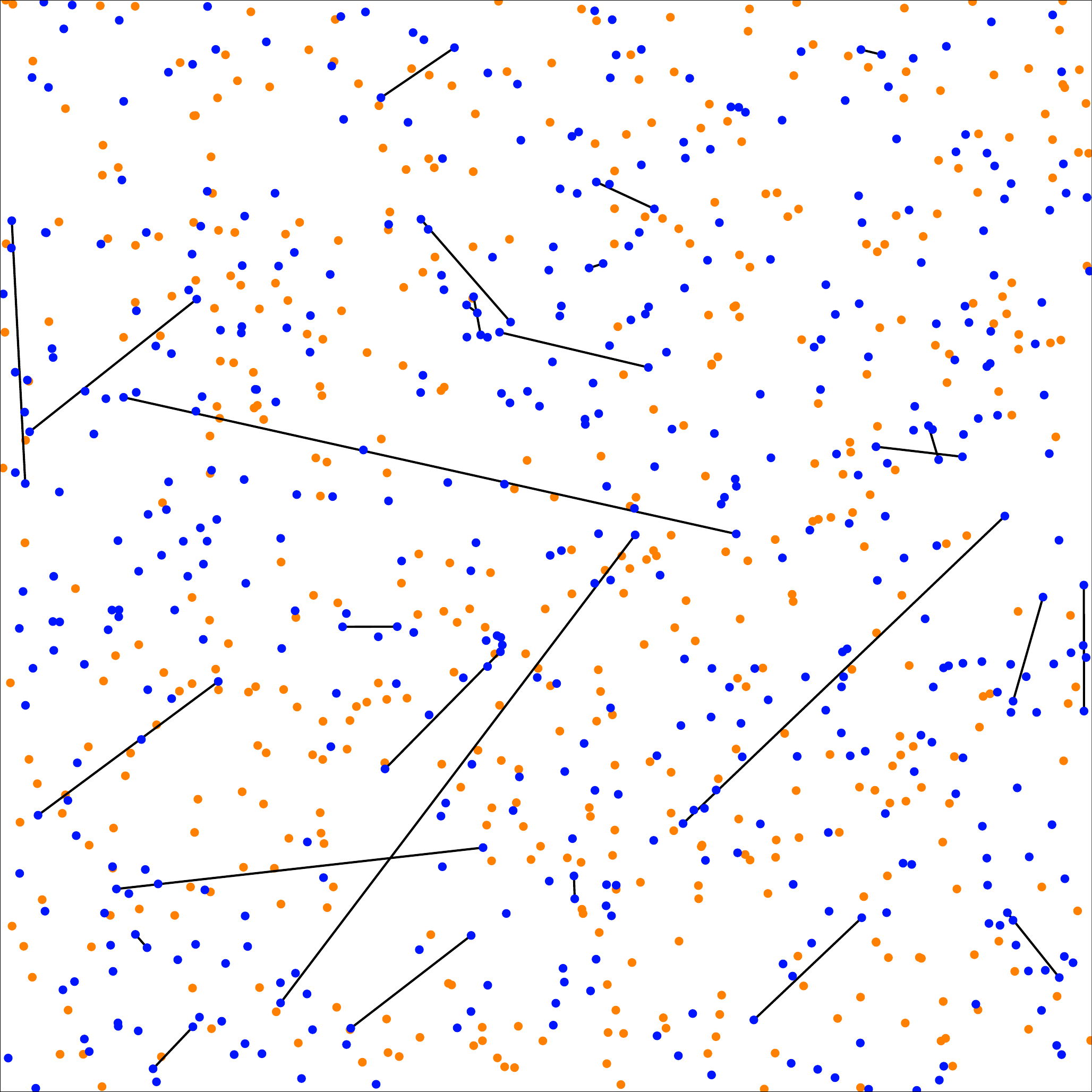}}
\caption{{\it Left}: All events which occured in an area of the sky map with dimensions 150 $\times$ 150 during the entire observation time. {\it Right}: The same events but identified as objects (blue) and random events (orange). Events which came from the same object are connected with a line. Less than 5\% of events come from objects which caused more than one event.}
\label{2dhidden}
\label{2dreveal}
\end{figure*}

	We can illustrate this difficulty by examining a small area of the sky map from our simulation. The left panel of Fig.~\ref{2dhidden} shows all the events that occurred in a 150 $\times$ 150 area during the entire 10 units of time. On the right the same events are shown but are identified as having come from objects (blue) or as random events (orange). Events which came from the same object are connected with a line. The difficulty of discovering moving objects by eye is evident. The 2-point function is statistically able to pick up on the rare occurrences where an object generates more than one event.

\subsection{Example 3: Moving sources, fixed sources, and a random component}	
	As a final example, a third class of objects are added to the simulation. These are stationary objects which do not move during the course of the observation. The presence of such objects should manifest itself as a spike in the 2-point function at $v=0$.
	
	The dimensions of the sky map are the same as in the two previous examples. The moving and stationary objects have the same event rate, $\l=0.01$, and spatial density, $n \approx 0.133$. The moving objects have the same velocity distribution as before. The random component has spacetime density $\r_0 \approx 0.00133$. Therefore, the total density of events is $\r = \r_0 + \r_1 + \r_2 = 0.004$, which is the same as in the last example. The subscript $2$ denotes the stationary objects. Each component contributes roughly the same number of events to the sky map.
	
	Including the stationary objects into the 2-point function just requires replacing the Rayleigh distribution with the Dirac delta function centered at $v=0$: $\pv(v) = \delta(v)$. Eq. \ref{Xvt2d} becomes,
\begin{equation}
\Xsvt= \frac{3 \,\r_2 \l \, (t_2-t_1) \delta_{v_1,0} }{\pi \, \rho^2 \,(v_2^2 -v_1^2)(t_2^3 - t_1^3)},
\label{Xsvt2d}
\end{equation}
and $\Xmvt$ is given by (\ref{Xvt2d}) except that the total density in the denominator includes the stationary objects, $\r = \r_0 + \r_1 + \r_2$. The function $\delta_{v_1,0}$ is $0$ if $v_1 > 0$ and is $1$ if $v_1=0$. The measured 2-point function $\Xvt$ is simply the sum of the 2-point functions for each class of objects: $\xi = \xi_1 + \xi_2$.

As before we choose $v_2 = v_1 + 1$ and $t_2 = t_1 + 1$. The results are plotted in the right panel of Fig.~\ref{fig:xi_MR_MRS}. The spike at $v=0$ due to the stationary objects is apparent. Its height is determined by both $\xi_1$ and $\xi_2$. Since the shape of $\xi$ when $v > 0$ can be measured the contribution of the moving objects to the spike at $v=0$ can be subtracted.

\section{Objects in 3 dimensions}
\label{sec:3d}

	The 2-dimensional situations examined so far are, of course, only toy models for astrophysical applications. In this section we develop a more realistic theory of the use of the 2-point function. The derivation of the form of $\xi$ is based on precisely the same arguments as in the 2-dimensional case. Simulations analogous to those in the previous section can also be performed in three dimensions and will agree with the theoretical form of $\xi$. In performing an actual measurement of $\xi$ simulations should be tailored to the specific application. We defer such detailed modeling to future work wherein we will apply the formalism to all-sky gamma-ray data \cite{GSK10b}.
			
	A diffuse emission all-sky map (e.g. from Fermi-LAT) is a 3-dimensional representation of a 4-dimensional process since we cannot measure line-of-sight distances for individual events. The distance to a source determines both its flux on Earth and its angular speed across the sky. This coupling between distance, speed, and flux is what makes the analysis more complicated. Previously, the velocity distribution $\pv(v)$ and the luminosity $\l$ were independent quantities. Now we must consider probability distributions which depend on both $v$ and $\l$: closer objects have higher angular speeds {\em and} look brighter than distant objects. While the spacetime 2-point function in this situation is still defined by (\ref{Xgeneral}) it is more difficult to derive the analogue of (\ref{XV2d}). The analysis of this section will develop the theory of the spacetime 2-point function in the case of a realistic sky survey.

\subsection{Summary of the measurement of $\xi$}

The computation of $\xi$ proceeds exactly as in the 2-dimensional case. The sky map consists of events, each having a directional coordinate (the apparent direction of the photon's origin) and a time coordinate. The ``distance'' between events is the angle between them measured along a great circle. The velocity of interest is now an angular velocity: the ``velocity separation'' of two events is defined as the angle between the two events divided by their time separation. The sky map is again a spacetime diagram, though not with the usual rectangular coordinates for the spatial axes. It can be visualized as a series of concentric spheres, each representing the celestial sphere, with different spheres corresponding to different slices of time (with $t$ increasing as the radius of the spheres increases). In this picture the worldlines of objects moving at constant angular speed are Archimedean spirals in spacetime. The volume of a region in this spacetime has units of solid angle $\times$ time.

In the next sections we derive expressions for $V(p)$ and $C(p;V)$, the latter in terms of parameters describing the various populations of objects which contribute to the sky map.

\subsection{The form of $V(p)$ in 3 dimensions}
One defines $V(p)$ as some volume of spacetime ${\cal S}$. Convenient choices include
\begin{eqnarray}
V(\T_1,\T_2;t_1,t_2)(p) &=& \{ p' \in {\cal S} :  t_1 \leq p'(t)-p(t) < t_2 \nonumber \\
&& \wedge \,\T_1 \leq d(p',p) < \T_2 \},
\label{V3drect}
\end{eqnarray}
where $p(t)$ is the time coordinate of the event $p$ and $d(p',p)$ is the angle between spacetime points $p$ and $p'$, and 

\begin{eqnarray}
V(\w_1, \w_2;t_1,t_2)(p) &=& \{ p' \in {\cal S } :  t_1 \leq p'(t)-p(t) < t_2 \nonumber\\
&& \wedge \, \w_1 \leq \frac{d(p',p)}{|p'(t)-p(t)|} < \w_2 \} , 
\label{V3dwt}
\end{eqnarray}
where $\w_1$ and $\w_2$ are angular speeds. 
These are the analogues of Eqs.~\ref{Vdefrect} \& \ref{Vdef}. In (\ref{V3drect}), $V(p)$ contains the events which occur in an annulus around $p$ with inner and outer radii $\T_1$ and $\T_2$ and which occur in the time interval $p(t)+t_1$ to $p(t)+t_2$. Note that when $t_1=0$ and $t_2=\infty$ this region is {\em exactly} that used for galaxy-galaxy correlation studies. In (\ref{V3dwt}), $V(p)$ represents all the events which could have been triggered by an object moving from $p$ if it had an angular speed between $\w_1$ and $\w_2$ and with the same time separation constraint.

If we are looking at a small area of the celestial sphere that can be approximated as flat space then we can choose $V(p)$ to be ``anisotropic''. i.e. choose volumes such as
\begin{eqnarray}
V(v_{x1},v_{x2};\,v_{y1},v_{y2}; \, t_1,t_2) &=& \{ p' \in {\cal S} :\\ 
&& t_1 \leq p'(t)-p(t) < t_2 ,  \nonumber  \\
&& \wedge \, v_{x1} \leq \frac{p'(x)-p(x)}{p'(t)-p(t)} < v_{x2}, \nonumber  \\ 
&& \wedge \, v_{y1} \leq \frac{p'(y)-p(y)}{p'(t)-p(t)} < v_{y2} \} \nonumber .
\label{V3dxyt}
\end{eqnarray}
This choice of $V(p)$ is useful when a class of moving objects has an anisotropic velocity distribution, or when the proper motion of the earth or the detector is important.

The volume of the region $V(p)$ is calculated in a way similar to the 2-dimensional case. For instance, the volume of the region specified by (\ref{V3drect}) is found by first computing the solid angle of the annulus between $\T_1$ and $\T_2$, 
and multiplying this by the time interval: 
\begin{equation}
V(\T_1,\T_2;\,t_1,t_2)=2\pi \left(\cos\T_1 - \cos\T_2\right)(t_2-t_1).
\label{Volrect}
\end{equation}
The volume specified in (\ref{V3dwt}) is slightly more complicated:

\begin{eqnarray}
V(\w_1, \w_2;  t_1, t_2)    &=& \int_{t_1}^{t_2} dt \int_0^{2\pi}d\phi \int_{\w_1 t}^{\w_2 t} \sin\T \, d\T  \nonumber \\
                            &=&  2\pi \left[\frac{ \sin(\w_1 t_2) - \sin(\w_1 t_1) }{\w_1}  \right] \nonumber \\
                           &-& 2 \pi \left[  \frac{ \sin (\w_2 t_2) - \sin (\w_2 t_1)}{\w_2} \right] .
\label{Volwt}
\end{eqnarray}
Equations~\ref{Volrect} and \ref{Volwt} hold only when $\T_2 < \pi$ and $\w_2 t_2 <\pi$, respectively. Otherwise the annulus begins to overlap itself. This is only an issue if one is searching for objects which moved across the entire sky during the observation period.

In the limit where $t_2 \rightarrow t_1 +dt$ and $\w_2 \rightarrow \w_1 +d\w$ (\ref{Volwt}) becomes (dropping subscripts)
\begin{equation}
V(\w_1, \w_2; \, t_1, t_2) \rightarrow dV(\w, t) = 2 \pi t \,{\sin}(\w t) \,d\w \, dt.
\label{dVoltwt}
\end{equation}
This is the analogue of the 2-dimensional (\ref{dV2d}).

As in the 2-dimensional case we now define a convenient coordinate system for every point on the celestial sphere. The coordinates $(\w, \phi, t)_p$ are related to the global celestial coordinates (plus time) as follows. First we consider a rotated set of spherical coordinates $(\capT, \Phi)_p$ in which $p$ is at the north pole and the line $\Phi=0$ intersects the north celestial pole. That is, the new and old coordinates are related by a rotation in which $p$ slides along a line of longitude to the north celestial pole. Then new coordinates $(\w, \phi, t)_p$ are related to $(\capT, \Phi)$ by $ \capT = \w t$ and $\Phi = \phi$.
This is a mapping from $(\w, \phi, t)_p$ to the global celestial coordinates (the time coordinate is unchanged). Using (\ref{dVoltwt}) we can write down the volume element in these coordinates. The spacetime volume (solid angle $\times$ time) between $\w$ and $\w+d\w$, between $\phi$ and $\phi+d\phi$, and between $t$ and $t+dt$ is
\begin{equation}
\label{dV3d}
dV_p(\w,\phi,t) = t \,{\sin}(\w t) \, d\w \, d\phi \, dt,
\end{equation}
and one can check that (\ref{Volwt}) is recovered as the integral 
\begin{displaymath} 
\int\limits_{\w_1}^{\w_2} \int\limits_0^{2 \pi} \int\limits_{t_1}^{t_2} dV_p(\w,\phi,t). \nonumber  
\end{displaymath}

\subsection{Ingredients needed to derive $C(p;V)$ in 3-D}
Besides the volume $V(p)$ we need to derive an expression for $C(p;V)$ in (\ref{Xgeneral}). This quantity depends on the properties of the sources which contribute events to the sky map. Generally, the sky is populated by different classes of objects, each with its own velocity distribution, luminosity function, and spatial distribution. Let's denote the different classes of objects by the subscript $i$. Then for each class we define the following functions.

\begin{itemize}

\item $\PiL(L)\,dL$ is the probability that an object of class $i$ has an intrinsic luminosity between $L$ and $L+dL$. $L$ is the number of photons per second emitted by the object. The distribution $\PiL(L)$ is normalized to 1: $\int_0^\infty \PiL(L)dL = 1$. The function $\PiL$ is commonly called the luminosity function of the population.

\item $n_i(R,\Ohat)$ is the physical number density of $i$-type objects which lie a distance $R$ away from the detector in the direction $\Ohat$ on the celestial sphere. This quantity has units [length]$^{-3}$.

\item $f_i(\vv;\, \Ohat)$ specifies the tangential velocity distribution of $i$-type objects. The quantity $f_i(\vv;\, \Ohat)d^2\vv$ is the probability than an object of class $i$ located in the direction $\Ohat$ on the celestial sphere has its tangential velocity vector in the range $d^2\vv$ around $\vv$. These velocities are proper velocities, measured relative to the earth (or the detector). ``Tangential'' means that the velocity is perpendicular to the line of sight\footnote{We have implicitly assumed that the line of sight velocity of any object is small enough that the change in its distance does not affect its flux. That is, the objects are all far enough away so that $\bar{v}_{\rm los} \bar{t}/R \ll 1$, where $\bar{t}$ is a measure of the time separation between the region $V(p)$ and the event $p$.}. This distribution is also normalized to 1. 

\item $\r$, defined above, is the average number of events detected per solid angle per time. It can be estimated from the sky map by dividing the total number of events by the time over which the sky map was measured and by the total solid angle of the map. For maps with large numbers of counts this estimator will be adequate. In practice, one may need to modify the procedure for surveys with unequal exposures across the sky: the quantity $\r$ may be position and time-dependent. If we divide $\r$ by the detector area $A$ we get $\tilde{\r}$, the total flux per solid angle.

\end{itemize}

\subsection{Derivation of $C(p;V)$ in 3 dimensions}
The quantity $C(p;V)$ is the probability of finding an event in the region $V(p)$ given that the detector reported the event $p$. It can also be thought of as the expected number of events in $V(p)$, given an event $p$. First we break $C(p;V)$ into the sum of 2 terms:
$C(p;V)$ = [the probability that the event $p$ was caused by an object which moved into the region $V(p)$ and triggered another event] + [the probability of finding an event in $V(p)$ for any other reason]. As in the 2-dimensional case the second term is simply $\r V(p)$.

The first term can be broken up into the product of 3 probabilities: [$C_1$: the probability that the event $p$ came from an object of class $i$ with luminosity $L$ located a distance $R$ from the detector] $\times$ [$C_2$: the probability that this object has a velocity that takes it into the region $V(p)$] $\times$ [$C_3$: the probability it triggers an event while in $V(p)$]. The product then needs to be integrated over $R$ and $L$ and summed over $i$.

The first factor, $C_1$, is the ratio of photons received from $i$-type objects with luminosity $L$ and distance $R$ in the direction $\Ohatp$ to the total number of photons received from the same direction:

\begin{equation}
\label{C1}
C_1 = \frac{1}{\r} \left[ n_i(R,\Ohatp)R^2 \, dR \right] \left[ \PiL(L)\, dL \right] \left[\frac{L\,A}{4 \pi R^2} \right],
\end{equation}
where $A$ is the effective area of the detector. It is worth noting that $C_1$ does not actually depend on $A$ since $\r$ will also be proportional to $A$.

The second factor, $C_2$, is the probability that an $i$-type object will have a velocity which takes it into the region $V(p)$. We will have to integrate over a range of velocities which correspond to the object moving into $V(p)$. It will, therefore, be useful to use the coordinate system defined in the discussion leading to (\ref{dV3d}). We can adapt the velocity distribution $f_i(\vv;\, \Ohat)$ to the new coordinates by introducing the function $f_i(v, \phi ;\, \Ohat)$ defined so that
\begin{equation}
f_i(v, \phi ;\, \Ohat) \, dv \, d\phi = f_i(\vv;\, \Ohat)d^2\vv.
\end{equation}
The quantity $f_i(v, \phi ;\, \Ohat) \, dv \, d\phi$ is to be interpreted as the probability that an object of type $i$ has tangential speed between $v$ and $v + dv$ and is moving in a direction between $\phi$ and $\phi+d\phi$, where $\phi$ refers to the coordinate label in our new coordinate system whose north pole coincides with the direction $\Ohat$ as described previously. Next we relate the distance to the object to its angular velocity using the relation $v = R \w$. Therefore, the quantity
\begin{equation}
f_i(R\w, \phi ;\, \Ohat) \, R \, d\w \, d\phi
\label{C2}
\end{equation}
gives the probability that the object, which is at a distance $R$, has angular speed between $\w$ and $\w +d\w$ and is moving in a direction between $\phi$ and $\phi+d\phi$. This expression is adapted for use with our new coordinate system.

The third factor, $C_3$, is the probability that the object triggers another event. This is simply given by

\begin{equation}
\label{C3}
C_3 = \frac{L \, A \, dt}{4 \pi R^2}.
\end{equation}
Combining this with (\ref{C2}) and integrating over $V(p)$ yields the quantity $C_2 \times C_3$:
\begin{equation}
\label{C2C3}
C_2 \times C_3 = \int\limits_{V(p)} f_i(R\w, \phi ;\, \Ohatp) \, R \frac{L \, A}{4 \pi R^2} d\w \, d\phi \, dt,
\end{equation}
which illustrates the benefits of our choice of coordinates $(\w, \phi, t)_p$. In words, $C_2 \times C_3$ is the probability that an $i$-type object with luminosity $L$, distance $R$, and starting at the location $\Ohatp$, moves into the region $V(p)$ and triggers an event. 

Now we can put together all the factors which make up $C(p;V)$ to find 
\begin{eqnarray}
C(p;V) &=&\r V(p) + \sum\limits_i \int\limits_{L=0}^{\infty} \, \int\limits_{R=0}^{\infty} C_1 \times C_2 \times C_3 \nonumber \\
	   &=& \r V(p) + \left(\frac{A}{4 \pi}\right)^2 \frac{1}{\r} \nonumber \\
	   &\times&  \sum\limits_i \int\limits_{L=0}^{\infty}\int\limits_{R=0}^{\infty}\int\limits_{V(p)}
	   \frac{n_i(R,\Ohatp)}{R} \, \nonumber \\
	  &\times&  \PiL(L) \,  L^2 \, f_i(R\w,\phi ;\, \Ohatp) \,d\w \, d\phi \, dt dL \, dR. 
\label{C3d}
\end{eqnarray}

In practice, since resolved objects will be removed from the sky map, the lower limit of the $R$ integral should be cut off so that these objects are not counted. If the detector can resolve any source with flux greater than $\Fres$ then the lower limit on the $R$ integral should be $\sqrt{L/4 \pi \Fres}$.

Of course, if $n_i(R,\Ohat)$ is cut off at a lower limit $R_{\rm min}$ and $\PiL(L)$ is cut off at an upper limit $L_{\rm max}$ such that $(L_{\rm max} / 4 \pi R_{\rm min}^2) < \Fres$ no changes need to be made to the limits of integration in (\ref{C3d}) since all $i$-type objects will be unresolved.

\subsection{The form of $\xi$ in 3 dimensions}
Finally, we can substitute (\ref{C3d}) into the definition of the 2-point function $\xi$ (\ref{Xgeneral}) and arrive at an expression for the 2-point spacetime correlation function in 3 dimensions, 
\begin{eqnarray}
\xi &=& \left(\frac{1}{4 \pi \tilde{\r}}\right)^2  \sum\limits_p \sum\limits_i \int\limits_{L=0}^{\infty}\int   \limits_{R=\sqrt{L/4\pi \Fres}}^{\infty}\int\limits_{V(p)} \frac{n_i(R,\Ohatp)}{R} \nonumber \\
&\times& \PiL(L) \, \,L^2 f_i(R\w, \phi ;\, \Ohatp) \,d\w \, d\phi \, dt \nonumber \\
&\times& \left[ \sum\limits_p V(p) \right]^{-1}.
%
%
%
\label{X3d}
\end{eqnarray}
Notice that the detector area $A$ has cancelled when using $\tilde{\r}$, the average flux per solid angle, instead of $\r$. It is also apparent that the contribution to the $\xi$ from different classes of objects as well as from objects of different distances and luminosities is additive. The observed 2-point function is simply the sum of contributions from different types of objects. As expected, the correlation is increased for brighter-appearing objects as is seen by the presence of $L^2$ and $R^{-1}$. The interplay between distance and angular speed appears in the argument of the velocity distribution $f_i$.

The expression for $\xi$ given by (\ref{X3d}) is a main result of this paper. In its general form, however, it is fairly opaque. We can get a qualitative feel for the 2-point function by calculating $\xi$ for a very simple model where we have only one class of objects. These objects have a constant number density $n$ and are found only at distances between $R_1$ and $R_2$. The intrinsic luminosity of all the objects will be fixed at $\l$ so that $\PiL(L) = \delta(L-\l)$. We choose an isotropic Maxwell-Boltzmann velocity distribution. Projected into 2 dimensions it becomes the Rayleigh distribution:
\begin{equation}
f(v, \phi ; \Ohat) \, dv \, d\phi = \frac{v}{a^2}e^{-v^2/2a^2}dv\frac{d\phi}{2\pi},
\end{equation}
independent of $\Ohat$. Finally we choose $V(p)$ to be given by (\ref{V3dwt}) in which $\phi$ runs from $0$ to $2\pi$. With these choices no quantity in (\ref{X3d}) depends on $p$ so both sums over $p$ disappear.

Let's look at the limiting form for $\xi$ by choosing an infinitesimal volume for $V(p)$ where $\w_2 = \w_1 + d\w$ and $t_2 = t_1 + dt$. Dropping the subscripts on $\w_1$ and $t_1$ we have the following expression for the 2-point function,
\begin{eqnarray} 
\xi(\w,t) &=& \frac{1}{2\pi t \sin(\w t)}\left(\frac{1}{4 \pi \tilde{\r}}\right)^2 n \l^2 \int\limits_{R_1}^{R_2} \frac{dR}{R}\frac{R\w}{a^2}e^{-(R\w)^2/2a^2} \nonumber \\
          &=& \frac{1}{2\pi t \sin(\w t)}\sqrt{\frac{\pi}{2}}\left(\frac{1}{4 \pi \tilde{\r}}\right)^2 \frac{n \l^2}{a} \nonumber \\
          &\times& \left[\Erf \left( \frac{R_2 \w}{\sqrt{2}a} \right) - \Erf \left(\frac{R_1 \w}{\sqrt{2}a} \right) \right].
\label{X3dtoy}
\end{eqnarray}

Note that the contribution to $\xi(\w,t)$ from objects at different distances serves to smear the influence of $f(v)$ so that $\xi$ is not simply proportional to the velocity distribution as it was in the 2-dimensional model. There is, however, a functional similarity to the 2-dimensinal case:
\begin{eqnarray}
\xi_{\rm 3D} &\sim& \frac{1}{V(p)} \frac{n \l^2}{\tilde{\r}^2} f( R \omega) \nonumber \\
\xi_{\rm 2D} &\sim& \frac{1}{V(p)} \frac{n \l^2}{\r^2} \pv, \, \, (\mathrm{using} \hspace{0.2cm} \r_1 = n\l \, \, \mathrm{in}\hspace{0.15cm} \mathrm{Eq.~} \ref{XdV2d})
\label{Xcompare}
\end{eqnarray}
where $f(R \omega)$ represents the smeared velocity distribution.

\subsection{Mock Fermi search for solar system bodies}

In order to verify the formulation of $\xi$ in (\ref{X3d}) we simulate a mock 5-year Fermi observation of nearby moving gamma-ray sources. These objects might correspond to a population of bodies in the asteroid belt (see Sec.~\ref{sec:conclusions} for motivation).

For simplicity, the detector is a stationary observer at the center of the solar system and the moving objects are placed on circular orbits with Keplerian velocities determined by their distance from the Sun ($v\propto R^{-1/2}$). The objects are distributed uniformly in a disk with uniform surface density between distances of 0.95 AU and 1.5 AU. For geometric simplicity, the inclination angles of the orbits are random so that the flux is statistically isotropic. Each object has the same luminosity and the closest object (at 0.95 AU) has a photon flux of $1.8\times10^{-10} \cm^{-2}\s^{-1}$. Note that this flux is below the point source detection limit of Fermi so that none of these moving objects would be individually identified as localized sources\footnote{In fact, moving sources will be more difficult to detect than stationary ones because of their apparent motion, i.e. standard point source analysis may be inefficient at detecting moving sources.}. The sky contains 7371 objects so that, for a 5 year Fermi observation (effective area $\sim 2000\,\cm^2$), the population of moving objects contributes roughly $2.5\times10^5$ events to the sky map. In addition, as in the 2-dimensional simulation, we include a population of stationary objects as well as completely random events. The stationary sources generate detected events at an average rate of 0.2 events per year and are distributed isotropically. The stationary and random components each comprise about $1.25\times10^5$ events so that the sky map contains about $5\times10^5$ events, 50\% from moving objects, 25\% from stationary sources, and 25\% random events.

\begin{figure}
\centering
\resizebox{3.3in}{!}
{\includegraphics{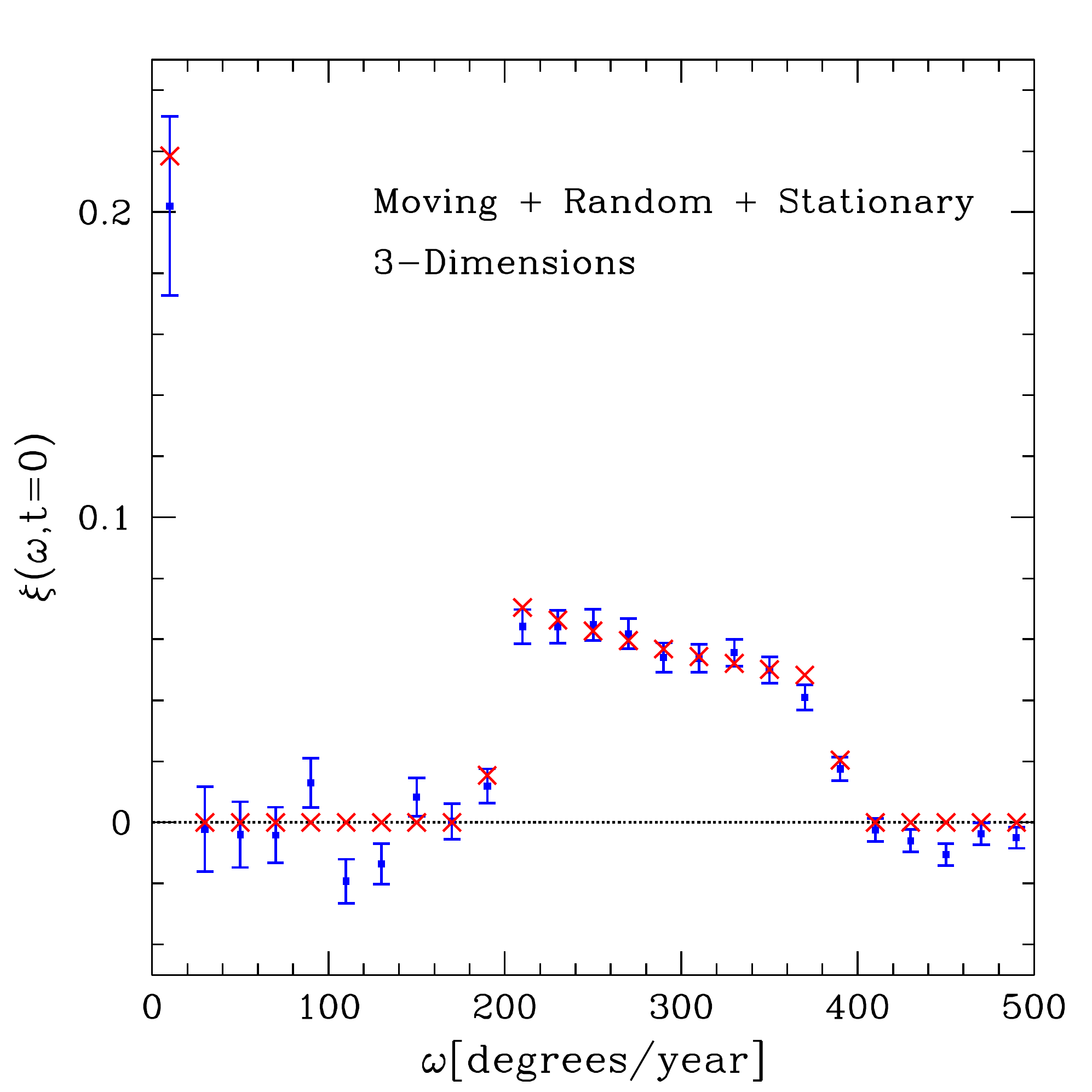}}
\caption{Results from a simulation of moving objects in the solar system, along with stationary sources and random noise. The correlation function is plotted for angular velocities between 0 and $500\deg$/yr. Red $\times$'s represent the theoretical value of $\xi$ calculated from (\ref{X3d}) while the blue squares show the measured value of $\xi$ from the sky map. The width of each angular velocity bin is $20\deg$/yr. Error bars are derived using (\ref{DX}). The spike at zero angular velocity is due to the presence stationary background sources. The correlation function is also non-zero between $\w=196\deg$/yr and $389\deg$/yr, corresponding to moving sources orbiting between $0.95$ and $1.5$ AU.}
\label{fig:3dsim}
\end{figure}

In computing the correlation function we use spacetime volumes $V(p)$ given by (\ref{V3dwt}) with $t_1=0$ and $\Delta T =t_2 -t_1=0.015$ yr ($\sim 5.5$ days). The angular velocity bins run from 0 to $500\deg$/yr in steps of $20\deg$/yr. Results of the measurement of $\xi$ for the different angular velocities are shown with blue squares in Fig.~\ref{fig:3dsim}. As with the 2-dimensional simulations, the error bars are found using (\ref{DX}). The presence of both the stationary and moving sources can be easily seen in the shape of the correlation function.

The theoretical value of $\xi$ based on the properties of the sources is a straightforward application of (\ref{X3d}). The number density of moving objects is
\begin{equation}
n(R)= \left\{ 
		     \begin{array}{lr}
		     	N_{\rm objs}/[{2\pi R(R_2^2-R_1^2)}] & R_1<R<R_2 \\
				0                                       & {\rm otherwise}.
			 \end{array}
		\right.
\end{equation}
Here, $R_1=0.95$ AU, $R_2=1.5$ AU, and $N_{\rm objs} = 7371$. The luminosity function and the velocity distribution are delta functions:
\begin{eqnarray}
P_L(L) &=& \delta\left(L - 4.45\times10^{17} \sec^{-1}\right) \\
f(v,\phi) &=& \frac{1}{2\pi} \,\delta\left(v - v_0 (R/R_0)^{-1/2}\right).
\end{eqnarray}
In the above, $R_0 = 1$ AU and $v_0 = 2\pi R_0/{\rm yr} \equiv \w_0 R_0$. The average event rate $\rho$ is estimated by dividing the total number of events by the solid angle of the sky map and by the observation time: $\rho=5\times10^5 /(4\pi \times 5 {\rm yr})$. Carrying through the calculation of (\ref{X3d}) yields $\xi$ for the particular choice of $V(p)$:
\begin{eqnarray}
\xi(\w_1,\w_2;\Delta T) &=& \frac{\Delta T}{4\pi \rho^2 V(\w_1,\w_2;\Delta T)} \left(\frac{LA}{4\pi R_0^2}\right)^2 \nonumber \\
                                  &\times&\frac{N_{\rm objs}}{R_2^2-R_1^2} \left(\frac{1}{R_a^2}-\frac{1}{R_b^2}\right).
\label{X3dsim}
\end{eqnarray}
The quantity $V(\w_1,\w_2;\Delta T)$ is the volume of the spacetime region given in (\ref{Volwt}), $R_a = {\rm Max}(R_1, R_0 (\w_0/\w_2)^{2/3})$, $R_b = {\rm Min}(R_2, R_0 (\w_0/\w_1)^{2/3})$, and $\xi=0$ if $R_a > R_b$. In (\ref{X3dsim}) the quantities $R_1,R_2,R_a$, and $R_b$ are in units of $R_0 = 1$ AU.

The correlation for the stationary objects is much simpler. It is equal to zero unless $\w_1=0$, in which case it is given by
\begin{equation}
\xi(\w_1=0, \w_2; \Delta T) = \frac{\Delta T \,N_{\rm stat} \,\lambda^2}{4\pi \rho^2 V(0,\w_2;\Delta T)},
\end{equation}
where $N_{\rm stat} = 1.25\times10^5$, the number of stationary objects and $\lambda = 0.2/{\rm yr}$, the detected event rate for each stationary object.

The total correlation function will be the sum of the correlation functions for each component. This sum is plotted in Figure \ref{fig:3dsim} as red $\times$'s, demonstrating that the formalism predicts the correct value for the correlation function.

Although intended as a toy model, this simulation captures the essential components of a large area analysis of Fermi data. In reality, Fermi has detected far more than $5\times10^5$ events. If all the components in our toy model were scaled up appropriately the detection of $\xi \neq 0$ would be even more significant.

\subsection{Errors and flux-limited vs. counts-limited surveys}

In three dimensions the errors on $\xi$ given by Eqs.~\ref{DX} \&  \ref{eq:SNR} also apply. As above, choosing the regions $V(p)$ requires balancing a large signal to noise ratio against having many independent choices of $V(p)$. In order to make more independent measurements of $\xi$ the size of $V(p)$ must decrease.

A larger $V(p)$ has its advantages and disadvantages. A large volume $V$ will decrease the fluctuations in $\xi$ because more events are collected in each such volume (the signal-to-noise contains a $V^{1/2}$ factor). On the other hand, having a lot of events in $V(p)$ which are uncorrelated to the event at $p$ will dilute the amplitude of $\xi$ because of the $\r V(p)$ term in the denominator (c.f. the definition of $\xi$ (\ref{Xgeneral})). There is then a tradeoff between the fluctuations in $\xi$ and the amplitude of $\xi$.

If the sky map has a large number of events then it is permissible to choose $V(p)$ to be small and still have small fluctuations in $\xi$. In the opposite limit, if the sky map is ``counts-limited'' then it will be necessary to choose $V(p)$ to have a large volume. The safest method for deciding is to run realistic simulations for various combinations of physical parameters and experiment with different choices for $V(p)$.

There is an additional requirement on $V(p)$ which depends on the detector's resolution. If one chooses $V(p)$ to be very small (in the angular sense) then one is essentially asking the detector to distinguish events at this angular scale. 
The detector has a smallest ``pixel size" and $V(p)$ cannot be smaller than that.

The most convenient choice for $V(p)$ when calculating $\xi$ according to (\ref{X3d}) is given by (\ref{V3dwt}). Unfortunately, this choice is inconvenient when dealing with a detector with a finite angular resolution (a real detector). The projection of the spacetime region $V(\w_1, \w_2; t_1$, $t_2)$ onto the celestial sphere must have an angular size no smaller than the detector's angular resolution. However, for fixed $\D\w = \w_2-\w_1$ and $\D t = t_2-t_1$, changing $\w_1$ and $t_1$ will change the projected angular size of $V(p)$. The bin sizes $\D \w$ and $\D t$ must be varied with $\w_1$ and $t_1$. An estimate of this constraint is that the angular resolution of the detector be no worse than $\T \approx \D(\w t) \approx \bar{\w} \D t + \D\w \bar{t}$, where $\D\w = \w_2-\w_1$ and $\D t = t_2-t_2$. 
	
All of these choices are part of the analysis, not the collection, of the data. If the diffuse background events are already in hand one can experiment with different choices for the $V(p)$'s to find the right balance between signal-to-noise and number of independent measurements of $\xi$ while maintaining the detector resolution constraint.  

Of course, it is best to use PSF information instead of an assumption of ``pixel size''. We have discussed this option for the 2-dimensional case (see (\ref{Cpsf})). The generalization to 3 dimensions is straightforward.

\section{Generalizations}	
\label{sec:discussion}	
There are several ways to make this technique more powerful. Here we mention two: the inclusion of spectral data and the use of $n$-point functions.

\subsection{Including spectral information}

	Not only do sky surveys keep track the direction and time of each photon they receive, they can also measure wavelength (or energy of the photon). The easiest way to make use of this information is to note that the above analysis holds for every wavelength separately. One can bin the events by energy, make separate sky maps for each energy bin, and then compute the 2-point function for each of the maps. 
	Typically, this procedure will add more data points than free parameters: the same distributions $n_i$ and $f_i$ are used for different energy bins. Only the luminosity functions will vary, though the physical parameters in $\PiL$ are likely to be universal over all energy bins. Thus, $\xi$ measured at one energy will be related to $\xi$ measured at another. As a result, an analysis which includes event energies can help in untangling the different components of the background.

\subsection{$n$-point functions}
	Another generalization of the 2-point function is, naturally, the $n$-point function. One asks, ``Given an event at $p$ what is the probability of finding events in $V_1(p)$ {\em and} in $V_2(p)$?'' If the objects move in straight lines then this probability will spike when $p$, $V_1(p)$, and $V_2(p)$ lie along a straight line. The jump from $n=2$ to $n=3$ is significant for this reason --- every pair of points is collinear but not every trio. The downside of measuring $n$-point functions (besides the computational cost) is that they require a much larger number of events to overcome statistical fluctuations. Recall that in our 2-dimensional toy model only 5\% of events came from objects which generated more than one event and that of these events, 95\% came from objects which generated exactly two events. Therefore, only $0.25\%$ of the events in the map came from objects which generated three or more events. Although they will be slightly more cumbersome, analytic forms for these higher correlation functions can be found by applying the same reasoning we used for the 2-point function.

\section{Discussion and Conclusions}
\label{sec:conclusions}

We present a new tool, based on the familiar 2-point correlation function, which can be applied to astrophysical maps of diffuse emission. The measured quantity $\xi$ is designed to detect the presence of moving objects, each of which is too dim to be resolved individually. We derived the form of $\xi$ based on the physical parameters which describe the classes of objects which might be present in the sky (\ref{X3d}). A measurement of $\xi$ along with the theoretical prediction for $\xi$ can be used to find best-fit quantities for the physical parameters describing the populations of objects. We emphasize that all the technology invented to study the angular 2-point correlation function can be directly applied to the generalization to the spacetime 2-point correlation function.

There are numerous applications of the derived formalism. An obvious place to start is the diffuse gamma-ray background measured by the Fermi-LAT instrument. The all-sky capabilities of LAT, coupled with its high angular resolution provide a convenient testbed where this technique can be applied. The interesting question is what kind of sources contribute to the gamma-ray background and also exhibit proper motion over the duration of observation. 

One potential source is the generation of gamma-rays from cosmic-ray interactions in rocky debris present in the solar system. Cosmic ray interactions with nuclei on a solar system body lead to hadronization, and the subsequent decay of neutral pions to a photon final state \citep{1970Ap&SS...6..377S,1981Ap&SS..76..213S,1986A&A...157..223D,2006PhRvD..74c4018K}. A detection of a large population of these sources is important as it provides information about the origin of the solar system and its evolution with time, as well as the energy spectrum and composition of the incident cosmic ray flux. 

The detection of gamma-rays from cosmic ray interactions with solar system bodies has been discussed in the context of past measurements by the Energetic Gamma Ray Experiment Telescope (EGRET) on board the Compton Gamma-ray Observatory, and measurements with Fermi-LAT \citep{2008ApJ...681.1708M,2009ApJ...692L..54M,2009arXiv0907.0541G}. Sources include small objects in the main asteroid belt, Trans-Neptunian objects in the Kuiper belt, as well as objects in the Oort cloud, including icy bodies such as comets. It was shown that for objects where the cosmic ray cascade fully develops (objects with size greater than $\sim 1$ m) it may be possible for Fermi to detect the cumulative gamma-ray emission from a collection of such bodies. These estimates are based on the distribution and composition of objects. Even though both of these quantities are partially constrained for objects in the main asteroid belt, large uncertainties are present for the populations in the Kuiper belt and the even more speculative Oort cloud.
It is conceivable that a very large number of bodies may be present in the outskirts of the solar system.

The proximity of these populations makes them ideal for an application of the spacetime correlation function, as each source will traverse an angular distance which is larger than the angular resolution limit of Fermi. Typical angular displacement (assuming Keplerian orbits) of an object at distance $d$ from the Sun is $\theta = 2 \pi \, {\mathrm{rad}} (\Delta T/\mathrm{yr}) (d / \mathrm{AU})^{-3/2}$ during the course of an integration for time $\Delta T$.  The composition of these objects can be assumed to be similar to the composition of the Moon, though their mass density varies considerably. This similarity in composition is convenient as the gamma-ray flux due to cosmic interactions with the lunar rock is well understood \citep{2007ApJ...670.1467M,1984JGR....8910685M} (see also \citep{2009arXiv0907.0543G,2009arXiv0912.3734G}). 
If we assume that the spectral shape of the gamma-ray emission from solar system bodies is similar to that of the rim of the Moon (emission above 600 MeV is dominated by the rim of the Moon rather than the lunar disc) and we scale the flux from the object to the flux from the Moon ($\Phi_M = 1.1 \times 10^{-6} \mathrm{cm}^{-2} \mathrm{s}^{-1}$, \cite{2009arXiv0907.0543G}), the flux from an object of radius $r$ at distance $d$ would then be $ \Phi = \Phi_M ( r / r_M) (d_M / d)^2$. For a distance to the Moon of $d_M=0.0024$ AU and a lunar radius of $r_M = 1740$ km, the total number of photons per year detected by the Fermi-LAT instrument (with an orbit-averaged effective area of 2000 $\mathrm{cm}^2$) is $\Phi \approx 2 \times 10^{-4} \mathrm{yr}^{-1} ( r / \mathrm{km} ) ( d / \mathrm{1AU} )^{-2}$.  Therefore, given this information, one can apply the spacetime correlation function to determine the abundance and radial distribution of solar system objects that contribute to the gamma-ray background \citep{GSK10b}.
It is important to note that even though a theoretical estimate of $\xi$ requires knowledge of the objects one is searching for, the measurement of $\xi$ requires no such knowledge.

Similar arguments can be used in search of the energetic neutrino signal from cosmic ray interactions with solar system bodies. The decay of kaons to charged pions leads to an energetic signal with a spectral signature that is different from the cosmic ray neutrino flux expected from spallation of nuclei. Therefore, energetic neutrinos from cosmic ray interactions with solar system bodies should be present in the signal measured by IceCube \citep{2006APh....26..155I}. The sources of these neutrinos will traverse an angular distance based on the distance of the source from the Sun, and therefore the spacetime correlation function derived here can be used in search of these sources. However, as in the case of gamma-rays, the uncertainties in the distribution and composition of small solar system bodies make predictions for such signal difficult. Nevertheless, a blind analysis of neutrino events from IceCube could place constraints on the parameters that describe the different populations of small bodies in the solar system. 

Another application is in the search for primordial black holes in the solar neighborhood. Primordial black holes may form in the early Universe through the collapse of large primordial fluctuations \citep{1971MNRAS.152...75H}. Current bounds on the abundance of such black holes are of order $\Omega_{\mathrm{PBH}} \sim 10^{-9}$ for most of the range of black hole masses \citep{2010ApJ...720L..67L}. If primordial black holes exist in an otherwise dark matter dominated Universe, they will acquire a dark matter halo \citep{Mack:2006gz,2007ApJ...662...53R}.  Dark matter annihilation around primordial black holes and/or high density ultracompact halos will result in gamma-ray emission \citep{2009ApJ...707..979R,2009PhRvL.103u1301S}.  Such objects with very small mass will in fact be very dense and survive in the Milky Way halo. If we assume that primordial black holes trace the distribution of dark matter in the Milky Way we can use their abundance to determine the angular distance that a black hole may traverse in a given time interval. For simplicity, let's assume that primordial black holes have mass $M_{\mathrm{PBH}} = 10^{-15} M_\odot$, $\Omega_{\mathrm{PBH}} = 10^{-9}$, and that the local dark matter density is 0.01 $M_\odot \mathrm{pc}^{-3}$. Then the mean distance between primordial black holes in the solar neighborhood is $\sim 10^{-2} \mathrm{pc}$. Assuming that this is the maximum distance to a primordial black hole, and that the mean velocity of primordial black holes is similar to the mean velocity of dark matter, i.e., 220 km/s, then the angular displacement of these gamma ray sources can be as large as 4.5 degrees in 10 years. As the angular resolution of Fermi is significantly less for energies greater than 1 GeV, constraints on the abundance and size of these black holes can be placed by applying the spacetime correlation function to the LAT all-sky map. 

A more speculative contribution to the gamma-ray background is from dark matter halos formed on scales close to the cutoff scale of the dark matter power spectrum. These objects typically have sub-solar masses \citep{1999PhRvD..59d3517S,2001PhRvD..64h3507H,2004MNRAS.353L..23G,2005JCAP...08..003G,2005PhRvD..71j3520L,2001PhRvD..64b1302C,2006PhRvL..97c1301P}. Even though their survival and abundance in the present-day Milky Way halo is unknown, it is possible that dark matter annihilation in these high-density objects may contribute to the gamma-ray background \citep{2008MNRAS.384.1627P,2008PhRvD..78j1301A}. The probability that such sources will exhibit spatial motion in the duration of the Fermi-LAT mission is directly linked to their abundance, and thus the use of the correlation function can provide information on the survival rate of these extremely early-forming objects. 

The spacetime correlation function can be applied to lensing surveys to search for compact objects in the Milky Way. Past studies suggest that up to 20\% of unseen matter is in the form of Massive Compact Halo Objects (MACHOs) \citep{2000ApJ...542..281A,2004ApJ...612..877U}. With the advent of dedicated surveys e.g., LSST, \citep{2009arXiv0912.0201L}, as well as astrometric missions such as SIM \citep{2008PASP..120...38U} and Gaia \citep{2008IAUS..248..217L}, it will be possible to generate time-domain maps of lensing events in dense stellar fields. Such information can be used to probe correlated events originating from the spatial translation of compact objects, thus probing the projected velocity distribution of the compact population in the Milky Way. In addition, it may also be possible to place constraints on the density, abundance and distribution of dark matter substructure \citep{2011ApJ...729...49E}.

Throughout the development of the analysis we assumed that the event rate due to any source was constant in time. There are many classes of astrophysical objects with time-dependent emission. Most notably, unresolved pulsars are thought to contribute to the diffuse gamma-ray background (e.g. \cite{2011ApJ...727..123W,2010JCAP...01..005F}). While these sources will not exhibit proper motion over the course of observations the temporal correlations of their emitted photons may be discovered through techniques based on the ones presented here \citep{2012MNRAS.421.1813G}. Essentially, one chooses the volumes $V(p)$ according to (\ref{Vdefrect}) (illustrated in the left panel of Fig.~\ref{fig:Vdef}), but with a non-trivial slicing along the time axis. Such a $V(p)$ picks up on stationary objects which exhibit correlations within their photon time series.

	The power of this analysis for untangling the contribution of different classes of sources requires that each class have ``different enough'' velocity, luminosity, and spatial distributions. For example, if two classes have similar velocity and spatial distributions then one may as well just treat them as a single class with a modified luminosity function. This points to a problem that is likely to be encountered in many realistic astrophysical applications: the angular velocities of almost all objects will be much too small to be resolved by a  detector. That is, when one combines the velocity distribution $f_i$ with the spatial distribution $n_i$ in (\ref{X3d}) it may be that $\xi=0$ at all angular velocities except in a tiny range near $\w=0$. This is because virtually all of the objects have distances and speeds such that their apparent proper motion is below the angular resolution of the detector. A large degeneracy is created and it will be impossible to pull out information about any specific class of objects. The fact that $\xi$ is not zero at $\w=0$ indicates the {\em existence} of objects. However, without being able to measure the shape of $\xi$ for different angular speeds $\w$ the 2-point function loses its value as a tool to untangle the contributions from different classes of objects.

	Of course, as the resolutions of detectors improve, the 2-point function becomes more useful. It is a straightforward task to calculate $\xi(\w_1,\w_2; t_1,t_2)$ for specific classes of objects and find out over what ranges of $\w$ and $t$ the correlation drops to zero. For example, if $\xi$ goes to zero around ($\w_1=\w', t_1=t'$) then a detector which has resolution better than $\T \approx \w' t'$ can measure the shape of $\xi(\w,t)$ as it goes from a maximum at ($\w_1=0,t_1=0$) to zero at ($\w_1=\w', t_1=t'$).

In summary, we introduced the spacetime correlation function, a statistical tool that can be used to search for the presence of moving, flux-unresolved sources in a diffuse background. This formalism has numerous applications. With large area sky surveys and long duration baselines the spacetime correlation function can be used to disentangle the contributions from spatially moving sources, and may aid in the discovery of new sources.

We thank the anonymous referee for comments and suggestions that improved the quality of this manuscript. We acknowledge useful conversations with John Beacom, Jacqueline Chen, Richard Cook, Ian Dell'Antonio, Scott Dodelson, Andrew Favaloro, Salman Habib, Arthur Kosowsky, David Laidlaw, Miguel Morales, Louie Strigari, Andrew Zentner. SMK and AGS are funded by NSF PHYS-0969853 and by Brown University. 

\bibliographystyle{mn2e}
\bibliography{manuscript}

\end{document}